\documentclass[a4paper,11pt]{article}
\pdfoutput=1 

\usepackage{jheparxiv} 

\usepackage[T1]{fontenc} 
\usepackage{dsfont}
\usepackage{tensor}
\usepackage{MnSymbol}
\usepackage{amsmath}
\usepackage{lipsum}
\usepackage{adjustbox}
\usepackage{amssymb}
\usepackage{subcaption}
\usepackage{mathrsfs}
\usepackage{graphicx}
\usepackage{mathtools}
\usepackage{stmaryrd}
\usepackage{twistor}

\usepackage{commath}
\usepackage{appendix}
\usepackage{yfonts}
\usepackage{tikz}
\usepackage{tikz-cd}
\usepackage{tkz-euclide}
\usepackage{capt-of}
\usepackage{xfp} 
\usepackage[outline]{contour} 
\usetikzlibrary{decorations.markings,decorations.pathmorphing}
\usetikzlibrary{angles,quotes} 
\usetikzlibrary{arrows.meta}
\usepackage{tikz-3dplot}
\usetikzlibrary{patterns}
\usetikzlibrary{math}
\usetikzlibrary{shadows.blur}
\contourlength{1.4pt}

\newcommand{\bPhi}{\boldsymbol{\Phi}}

\newcommand{\veps}{\varepsilon}

\newcommand{\dbl}{\Dot{\beta}}

\newcommand{\dbeta}{\Dot{\beta}}

\newcommand{\m}{m}

\tikzset{>=latex} 
\colorlet{myred}{red!80!black}
\colorlet{myblue}{blue!80!black}
\colorlet{mygreen}{green!80!black}
\colorlet{mydarkred}{red!50!black}
\colorlet{mydarkblue}{blue!50!black}
\colorlet{mylightblue}{mydarkblue!6}
\colorlet{mypurple}{blue!40!red!80!black}
\colorlet{mydarkpurple}{blue!40!red!50!black}
\colorlet{mylightpurple}{mydarkpurple!80!red!6}
\colorlet{myorange}{orange!40!yellow!95!black}

\tikzstyle{world line}=[myblue!60,line width=0.4]
\tikzstyle{world line t}=[mypurple!60,line width=0.4]

\title{A systematic approach to celestial holography: a case study in Einstein gravity}

\author[a]{Wei Bu,}\emailAdd{w.bu@sms.ed.ac.uk} 
\author[b]{Sean Seet}\emailAdd{sxes2@cam.ac.uk} 

\affiliation[{a}]{School of Mathematics and Maxwell Institute for Mathematical Sciences\\
University of Edinburgh, EH9 3FD, UK}

\affiliation[b]{Department of Applied Maths \& Theoretical Physics,\\
Wilberforce Road, Cambridge CB3 0WA,\\
United Kingdom}
\abstract{We propose a systematic approach to celestial holography in massless theories beginning by studying the implications of properly incorporating field configurations built using the eigenstates of central interest: massless conformal primary wavefunctions that diagonalize the dilatation generator. Due to their singular behaviour on the locus $k\cdot x=0$, they do not belong to the space of Fourier decomposable functions, and incorporating them in the path integral domain requires careful manipulations. In this paper, we include these singular field configurations by a splitting procedure using large pure gauge/diffeomorphism transformations on the action functional. We demonstrate that doing so splits the action into an integrand supported on the singular locus $k\cdot x=0$ and an integrand on the rest of the space. Mellin transforms single out the scalings/conformal dimension in $x$, geometrically, we treat this as a proper non-compact scaling reduction, where we are able to further isolate the dynamics of the large pure diffeomorphism transformations. This takes the form of 2d chiral CFT on a 2d sphere on the singular locus $k\cdot x=0$ - the celestial sphere where the null cone of the origin cuts $\scri$. Using this framework, we study Einstein gravity perturbatively around its self-dual sector, where the resulting microscopic 2d CFT couples to bulk scattering states. We are able to obtain an explicit representation of the $\cL w_{1+\infty}$ algebra and leading soft splitting functions. With further marginal deformations, we also write down effective interaction vertices which provide form factors of tree-level graviton scattering in Minkowski space.}
\begin{document} 
\maketitle
\flushbottom
\section{Introduction}
Large gauge/diffeomorphism symmetries of massless gauge/Einstein gravity theory in 4d Minkowski space are additional symmetry transformations of the theory that have nonzero action on field configurations on the conformal boundary $\scri$. As genuine symmetries of the theory, their Ward identities were found to be equivalent to soft theorems in gauge/gravity tree level scattering amplitudes \cite{Kapec:2014zla,Gabai:2016kuf,Campiglia:2016hvg,Campiglia:2016efb}. 
This highlights the fact that large gauge transformations are not merely gauge artifacts, but affect the dynamics of bulk fields. 

Explicit forms of the generators of these symmetries have been derived by expanding the potential/metric in a large $r$ limit in radial gauge, giving pure gauge/diffeomorphism profiles appearing as the leading $1/r$ contributions in the large $r$ limit \cite{Strominger:2017zoo,Duary:2022onm,Donnay:2018neh,Pasterski:2021fjn,Donnay:2020guq}. Intuitively, although they are pure gauge and naively can be gauged away, their non-vanishing at large $r$ means that they should be thought of as asymptotic shift symmetries that move the field configurations in the space of vacua.

Despite its unconstrained nature, such free asymptotic data moves the bulk theory in the allowed space of boundary conditions, which in turn puts restrictions on the allowed bulk kinematics and dynamics. Its Ward identities seem to have knowledge about bulk scattering with soft scattering state insertions. Motivated by this, we propose to study the following questions: 

\textit{Could the large pure diffeomorphism profiles have independent dynamics at asymptotic infinity? Is there a holographic way of interpreting the knowledge the boundary system has of bulk dynamics in Einstein gravity?}

In particular, instead of starting from bulk dynamical observables by Mellin transforming known S-matrices in 4d Minkowski space, we make an attempt to answer this question at the level of action functionals in the path integral. The key observation is that from a path integral perspective, if one wants to investigate the extra field configurations provided by large pure diffeomorphisms (which is the case of interest in celestial holography), one is required to sum over such field configurations. We include these extra field configurations by promoting large pure diffeomorphism profiles to quantum fields and performing a path integral over them. These quantum fields enter the Lagrangian as pure diffeomorphisms with scaling weight $0$ in $x^\mu$ such that they do not fall off at infinity. Hence one is free to choose horribly non-normalizable basis on $\scri$ to expand such fields. One particular choice of interest is the conformal primary eigenstate \cite{Pasterski:2016qvg,Pasterski:2017kqt} which are eigenstates of dilatation and singlets under the little group of $SO(3,1)$. Such eigenstates can be obtained by taking the Mellin transform of the usual massless plane wave wavefunctions in terms of their energy scale, which gives eigenstates of the form $1/(k\cdot x)^\Delta$ with $k^\mu$ a (normalised) null 4-vector and generic $\Delta\in\mathbb{C}$. Note that for $\Delta>0$, they are not smooth or (generically) square normalizable, and therefore fall outside the standard path integral domain of Fourier decomposable field configurations. We demonstrate that our "splitting" procedure for including these non-square normalizable singular field configurations that results in a theory that includes the original theory as a subsector, but also gives us the language to analyse the dynamical interplay between large gauge transformations and bulk processes in the meantime.

An interesting feature is that the action functional splits into two patches of spacetime where the singular bulk configurations live on the patch where $k\cdot x\neq 0$, while the pure diffeomorphism profiles live on the patch where $k\cdot x=0$.

We have employed this philosophy in the context of 4d gauge theory in \cite{Bu:2023cef,Bu:2023vjt} with the idea of holographic reduction and mechanisms of twistor geometry. Twistor geometry emerges in a natural way to properly describe the codimension-1 Minkowski locus $k\cdot x=0$. For a generic null vector $k^\mu\propto \lambda^{\alpha}\bar\lambda^{\dal}$, regardless of its magnitude, one could consider the projective null cone $\mathbb{CP}^1_{\lambda,\bar\lambda}$ fibring over Minkowski space, with appropriate complexification, we consider the uplift of the spacetime theory to twistor space $\mathbb{PT}$. With the two patches $k\cdot x=0$ and $k\cdot x\neq 0$ clearly labelled on $\mathbb{PT}$, we can further isolate the piece of the action living on the 4-dimensional patch $k\cdot x=0$, with topology $S^2\times\mathbb{R}^2$ by reducing the action functionals two real dimensions down to a 2-dimensional conformal theory on $S^2$ with all fields labelled by mode numbers along $\mathbb{R}^2$. This is the geometric way of picking out the scaling in $|x|$, which is precisely the function of the Mellin transform. Indeed one can see this as geometrically performing Mellin transform on off-shell fields. The chiral 2d CFT we recover contains independent dynamics with couplings and interactions inherited from the bulk theory before splitting\footnote{This is reminiscent of holographic reduction \cite{Sleight:2023ojm,Cheung:2016iub,deBoer:2003vf} and the edge mode separation \cite{Kim:2023qbl}.}. We further study it in detail, obtaining results as summarized below.

\paragraph{Summary of results:}
In this paper, we extend the formalism to firstly self-dual sector of 4d Einstein gravity, and consider further deformations of the 2d CFT we obtain to compute form factors matching full Einstein gravity. Using the twistor action for self-dual gravity, we give a first principle derivation of the 2d CFT guided by this principle (we have suppressed the labels on individual fields for clarity):
\begin{multline}\label{intro_2dcft}
    S=\sum_{p=1}^{\infty}\int_{\mathbb{Z}+\im\mathbb{R}}\d^2\Delta\int_{\mathbb{CP}^1_\lambda\times\mathbb{CP}^1_\mu}\D\lambda\D\mu \wedge \bar\delta^{(p-1)}([\mu \bar\lambda])q^{p+1}\la\lambda\hat\lambda\ra^p\\
    \left[\tilde \phi \bar\partial \phi + \eta\bar\partial\tilde\eta +h^{M'} \underbrace{\left(\{ \tilde\phi,\phi\}+\{\tilde\eta,\eta\}-\Delta\tilde\phi\right)}_{J_{\Delta,p}}+\tilde h^{M'}\underbrace{\left(\{\tilde\eta,\phi\}-(\Delta-2\im)\tilde\eta\right)}_{\tilde J_{\Delta, p}}\right]\,.
\end{multline}
Although the integral is over some 2 complex dimensional space, the holomorphic delta function $\bar\delta^{(p-1)}([\mu \bar\lambda])$ is restricting us to a 2d CFT on formal neighborhoods of $\mathbb{CP}^1_{[\mu \bar\lambda]=0}$. The sum over $p\in\mathbb{Z}_+$ labels the localisation to the $(p-1)$th order formal neighborhood of the locus $\mathbb{CP}^1_{[\mu\bar\lambda]=0}$, $\Delta\in\mathbb{Z}+\im\mathbb{R}$ denotes the mode number along $\mathbb{R}^2$. $\phi$, $\eta$ normalised boundary values of certain bulk fields and $\tilde\phi$, $\tilde\eta$ are the free pure diffeomorphism modes. A natural coupling term emerges between bulk gravitons $h^{M'}$, $\tilde h^{M'}$ and combinations of the free chiral fields which we named the currents.  
Using this Lagrangian formulation, we are able to isolate the boundary currents that couple to the bulk gravitons, whose OPEs reproduce the $\cL w_{1+\infty}$-algebra \cite{Strominger:2021mtt,Guevara:2021abz}\footnote{We note that this has been obtained using Koszul duality in \cite{Costello:2022_CP,Costello:2023hmi} and directly from twistor actions \cite{Adamo:2021lrv,Mason:2022hly,Bittleston:2024rqe}.}
\begin{equation}
    J_{\Delta,p-1}(\lambda_1)J_{\Delta',p'-1}(\lambda_2)\sim\frac{\Delta (p'-1)-\Delta'(p-1)}{\la\lambda_1\lambda_2\ra}\,J_{\Delta+\Delta',p+p'-2}(\lambda_2)\,,
\end{equation}
without the wedge condition on the mode numbers $\Delta\in\mathbb{Z}+\im\mathbb{R}$ and $p\in\mathbb{Z}_+$.
We also recover the gravity splitting function \cite{Bern:1998sv,Nguyen:2009jk} from a study of currents coupling to usual Fourier modes of graviton scatterings states:
\begin{equation}\label{intro_OPEeq}
J(\lambda_1,p_1^{\alpha\dal})J(\lambda_2,p_2^{\alpha\dal})\sim \frac{1}{2}\,\frac{[12]\,\D\lambda_2}{\la 12\ra\la\lambda_1\hat{\lambda}_1\ra\la\lambda_2\hat{\lambda}_2\ra}\frac{\la\lambda_1\iota\ra\la\lambda_1\upsilon\ra}{\la\lambda_2\iota\ra\la\lambda_2\upsilon\ra}\,J(\lambda_1,p_1^{\alpha\dal}+p_2^{\alpha\dal})+1\leftrightarrow 2\,,
\end{equation}
where we have incorporated arbitrary reference spinors $\iota^\alpha$ and $\upsilon^\alpha$ and some massless bulk Fourier mode $p_i^{\alpha\dal}$ with $p_i^2=0$. We further add marginal deformations to the 2d chiral CFT \eqref{intro_2dcft} to provide effective background interaction vertex. Together with the OPEs for the currents coupling to scattering states \eqref{intro_OPEeq}, we are able to compute all-multiplicity MHV correlation functions which match form factors in full Einstein gravity\footnote{We note that this computation is purely 2d, which differs from usual twistor space bulk computations of scattering amplitudes \cite{Adamo:2013tja,Adamo:2011cb}.}.  

In section \ref{sec:GR_twistor}, we review the basics of the geometry of twistor space and various formulations of the action for the self-dual sector of Einstein gravity together with their extension to full general relativity. In section \ref{sec:review_reduction}, we explain the general philosophy of formally splitting the theory by allowing certain pure gauge transformations to introduce bulk field configurations expandable in conformal primary basis. As a demonstration of the formalism to isolate independent dynamics of the pure gauge profiles, we explicitly write down the splitting procedure in the simple case of a free scalar. We hope the reader finds this part of the discussion easy to follow. In section \ref{sec:reduction_2d}, guided by the same philosophy, we explicitly work out the formal splitting in self dual gravity and derive the 2d CFT living on the special locus $\mathbb{CP}^1_{[\mu\bar\lambda]=0}$. Given the 2d CFT, we further discuss two perspectives one could take to view the couplings of the CFT fields with bulk graviton modes (scattering modes or conformal primary modes). This allows us to read off currents composites of the aforementioned CFT fields whose interaction is inherited from the bulk theory. In section \ref{sec:OPEs}, we compute the operator product expansion (OPE) of these currents and find concrete realisations of the $\cL w_{1+\infty}$ algebra and leading soft splitting function. In section \ref{sec:deformation}, we add further marginal deformations to the 2d CFT, which takes the form of a background interaction vertex. This sources additional effective interactions in the split theory and allows us to completely decouple the bulk and boundary dynamics. We present the formalism to compute all tree-level MHV type correlation functions in the presence of such an effective vertex and match the results to form factors of bulk graviton scattering. In the appendices, we further include discussions which otherwise interrupts the flow of the main text. In appendix \ref{Appendix:C}, we explain how to recover form factors from marginal deformations of the 2d chiral action. In appendix \ref{sec:charges}, we compute the symmetry charge for the large pure diffeomorphism of our twistor action functional, which localises to a sphere with appropriate choices of variations in the symplectic form on phase space. Such charges are found to shift the large pure diffeomorphism profiles and bestow them with higher singularity behaviours along $\scri$. 

\section{Review of twistor action for general relativity}\label{sec:GR_twistor}
In this section we discuss the required twistor background knowledge. The reader interested in the motivation and seeing a simple example of our procedure is encouraged to skip to section \ref{sec:review_reduction}.
\subsection{Twistor space}\label{subsec:PT}
The projective twistor space $\mathbb{PT} = \mathbb{R}^4 \times \mathbb{CP}^1$ is the space of complex structures over $\mathbb{R}^4$. It can be understood as the removal of a $\mathbb{CP}^1$ from $\mathbb{CP}^3$ \cite{Penrose_Rindler_1988}:
\begin{equation}
    \mathbb{CP}^3 = S^4 \times \mathbb{CP}^1 \xrightarrow[]{\text{remove a point on $S^4$}} \mathbb{R}^4 \times \mathbb{CP}^1 = \mathbb{PT}\,,
\end{equation}
and is therefore a complex 3-manifold, inheriting the canonical complex structure on $\mathbb{CP}^3$. Throughout this paper, we will work with $\mathbb{PT}^\cO$, in which we further remove the origin of $\mathbb{R}^4$. This removes the locus $\{0\}\times \mathbb{CP}^1$.
\begin{equation}
    \mathbb{PT}^\cO \cong (\mathbb{R}^4\setminus\{0\}) \times \mathbb{CP}^1\,.
\end{equation}
$\mathbb{PT}^\cO$ is coordinatised by $(x^{\alpha \dot \alpha}, \lambda_{\beta})$, in which $x^{\alpha \dot \alpha} \in \mathbb{R}^4 \setminus\{0\}$ and $\lambda_{\beta}$ are homogeneous coordinates on $\mathbb{CP}^1$. We will often use $\mathbb{CP}^3$ homogeneous coordinates $Z^A=(\mu^{\dot \alpha}, \lambda_{\alpha})$ with $A=0,1,\Dot{0},\Dot{1}$ being $SL(4)_{\mathbb{C}}$ indices and $\alpha=0,1$, $\dal=\Dot{0},\Dot{1}$ being $SU(2)$ indices. The coordinates are related by:
\begin{equation}\label{incidence_relation}
    \mu^{\dot \alpha} = x^{\alpha \dot \alpha} \lambda_{\alpha} \implies x^{\alpha \dot \alpha} = \frac{\hat\mu^{\dot \alpha}\lambda^{\alpha}-\mu^{\dot \alpha}\hat \lambda^{\alpha}}{\la \lambda \hat \lambda \ra}\,.
\end{equation}
where we have defined Euclidean conjugation:
\begin{equation}
    \hat \lambda^{\alpha} = \begin{pmatrix}
        -\bar \lambda^1 \\ \bar \lambda^0
    \end{pmatrix}, \quad \hat{\mu}^{\dal} = \begin{pmatrix}
        -\bar \mu^{\Dot{1}} \\ \bar \mu^{\Dot{0}}
    \end{pmatrix}\,.
\end{equation}
We also use the shorthand notation to denote spinor helicity contractions:
\begin{equation}
    \la a b \ra := a^\alpha b_{\alpha} = \veps^{\alpha \beta} a_{\beta}b_{\alpha}, \quad [ab]:= a^{\dot \alpha} b_{\dot \alpha} = \veps^{\dot \alpha \dot \beta} a_{\dot \beta}b_{\dot \alpha}\,,
\end{equation}
where $\veps^{\alpha\beta}$ and $\veps^{\dal\dbeta}$ are the $SL(2,\mathbb{C})$ Levi-Civita symbols. 
In terms of $\mu^{\dot \alpha}$ coordinates, we write $\mathbb{PT}^\cO$ as
\begin{equation}
    \mathbb{PT}^\cO :=\left\{(\mu^{\dal},\lambda_{\alpha})\in\mathbb{CP}^3\vert \lambda_\alpha\neq0 \,\text{ and }\, \mu^{\dal}\neq 0\right\}\,.
\end{equation}
\subsection{Self dual gravity on spacetime}\label{subsec:SDGR}
The setting for this discussion is 4d spacetime with vanishing cosmological constant. In 4 spacetime dimensions, a self-dual (resp. anti-self dual) Einstein spacetime is a "half-flat" solution to Einstein's equations that has the undotted (resp. dotted) Weyl spinor vanishing.
\begin{equation}
    \Psi_{\alpha \beta \gamma \delta} = 0\,.
\end{equation}
In Lorentzian signature, the dotted and undotted Weyl spinors are not independent but complex conjugates,
    \begin{equation}
        \overline{\left(\Psi_{\alpha \beta \gamma \delta}\right)} = \bar {\Psi}_{\dot\alpha \dot\beta \dot\gamma \dot\delta}\,,
    \end{equation}
so the "half-flatness" condition implies the vanishing of the full Weyl curvature, and therefore that the spacetime is conformally flat. To get (anti-)self-dual Einstein metrics with nonvanishing Weyl curvature, we consider Euclidean, split, or complexified spacetimes. There exist several action formulations for self-dual gravity. Chalmers and Siegel \cite{Chalmers:1996rq} proposed the following action (in rather different language) in terms of the vierbein $e^{\alpha \dot \alpha}$ and the symmetric undotted spin bundle connection $\Gamma^{\alpha \beta}=\Gamma^{(\alpha \beta)}$ 
\begin{equation}
        \int_{\mathbb{R}^4} \Sigma(e)_{\alpha \beta} \wedge \d \Gamma^{\alpha \beta}, \quad \Sigma(e)_{\alpha \beta} := e_{(\alpha}{}^{\dot \alpha} \wedge e_{\beta) \dot \alpha} \,.
\end{equation}
Capovilla, Dell, Jacobson and Mason \cite{Capovilla:1991qb} proposed an action in terms of a triplet of 2-forms $\Sigma_{\alpha \beta} = \Sigma_{(\alpha \beta)}$ and a Lagrange multiplier $\psi^{\alpha \beta \gamma \delta} = \psi^{(\alpha \beta \gamma \delta)}$
\begin{equation}\label{CDJM_SDG}
        \int_{\mathbb{R}^4} \Sigma_{\alpha \beta}\wedge \d \Gamma^{\alpha \beta} + \psi^{\alpha \beta \gamma \delta} \Sigma_{\alpha \beta} \wedge \Sigma_{\gamma \delta}\,,
\end{equation}
in which the Urbantke metric \cite{Urbantke_1984}
\begin{equation}
        \sqrt{g} g_{\mu \nu}(\Sigma) = \epsilon^{\sigma \rho \tau\theta} (\Sigma^{\alpha \beta})_{\mu \sigma}(\Sigma^{\beta}_{\gamma})_{\rho \tau}(\Sigma^{\gamma \alpha})_{\theta \nu}\,,
\end{equation}
for $\Sigma$ is self-dual on the support of the field equations for the Lagrange multipliers $\Gamma, \psi$. The Capovilla-Dell-Jacobson-Mason action reduces to the Chalmers-Siegel action by integrating out $\psi^{\alpha \beta \gamma \delta}$ and learning that the 2-form $\Sigma_{\alpha \beta}$ is simple. Capovilla, Dell and Jacobson \cite{Capovilla:1991kx} proposed the following action in terms of the linearised undotted Weyl spinor $\psi$ and the undotted spin bundle connection $q^{\alpha \beta}$,
\begin{equation}\label{CDJ_SDG}
        \int_{\mathbb{R}^4}\psi^{\alpha \beta \gamma \delta} \d q_{\alpha \beta} \wedge \d q_{\gamma \delta}\,,
\end{equation}
in which the Urbantke metric for $\d q$ is self-dual on the support of the field equations. It was shown in \cite{Bittleston:2022nfr} that the action \eqref{CDJ_SDG} is related to the action \eqref{CDJM_SDG} by the substitution $\Sigma_{\alpha \beta} = \d q_{\alpha \beta}$ and enforcing the $\d$-closure of $\Sigma$ with a Lagrange multiplier.
    
From Penrose's nonlinear graviton construction, it is known that a complex structure on twistor space corresponds to the conformal class of a metric describing a self-dual Einstein background. As could have been anticipated from the diversity of formulations for a spacetime action, various action formulations exist on twistor space of "BF" form that impose the vanishing of the Nijenhuis tensor (and therefore the existence of a complex structure) on twistor space \cite{Mason:2007ct, Sharma:2021gcz}. These have been shown to be classically equivalent to various spacetime actions for self-dual gravity. Here we will swiftly recap the twistor action most closely related to the Capovilla-Dell-Jacobson (CDJ) action for self-dual gravity, presented in terms of perturbations around a flat background.
\subsection{Twistor action for CDJ formulation}
On spacetime with vanishing cosmological constant, for small perturbations around flat space we preserve the fibration of twistor space over the $\mathbb{CP}^1_{\lambda}$. The fundamental fields of the theory are $h \in \Omega^{0,1}(\mathbb{PT},\mathcal{O}(2))$, $\tilde h \in \Omega^{0,1}(\mathbb{PT},\mathcal{O}(-6))$. We are supplied with the simple bitwistor $I^{AB}$ that breaks conformal invariance. The action is \cite{Penrose:1985bww,Penrose:1986ca,Mason:2007ct}:
    \begin{equation}\label{hth_twistor_action}
       S_{\text{SD}}= \int_{\mathbb{PT}^\cO} \D^3Z \wedge \tilde h \wedge \left(\bar \partial h + \frac{1}{2}\{h,h\}\right)\,.
    \end{equation}
The almost complex structure is defined by the $\bar\nabla$ constructed as a Hamiltonian deformation of the flat space $\bar \partial$ operator with respect to the symplectic form:
    \begin{equation}\label{poisson_bracket}
        \bar\nabla:=\bar \partial + \{h,\cdot\}, \quad \{f,g\}:=I^{AB}\frac{\partial f}{\partial Z^A}\frac{\partial g}{\partial Z^B}=\epsilon^{\dot \alpha \dot \beta}\frac{\partial f}{\partial \mu^{\dot \alpha}} \frac{\partial g}{\partial \mu^{\dot \beta}}\,.
    \end{equation}
where $I^{AB}=\frac{1}{2}\begin{pmatrix}
\veps^{\dal\dbeta} & 0\\
   0 & 0
\end{pmatrix}$ is the upstairs infinity twistor.
The equation of motion of $\tilde h$ enforces the vanishing of the Nijenhuis tensor $\mathcal{N}$ of the deformed complex structure, being the obstruction to the nilpotence of $\bar\nabla$.
A point in $\mathbb{R}^4$ corresponds to a holomorphic twistor line in the complex structure defined by $\bar\nabla$. The defining equation for such a twistor line is
    \begin{equation}
        L_x := \{\mu^{\dot \alpha}-F^{\dot \alpha}(x,Z^A)=0\}, \quad \bar\nabla\left(\mu^{\dot \alpha}-F^{\dot \alpha}(x,Z^A)\right)=0\,,
    \end{equation}
where $F^{\dot \alpha}(x)$ is a smooth function of $Z^A$ of weight $\mathcal{O}(1)$ parameterised by $x^{\alpha \dot \alpha}$ that solves the above holomorphy equation. Note that when $h=0$, $F^{\dot \alpha}(x,\lambda)=x^{\alpha \dot \alpha} \lambda_{\alpha}$, we recover the flat twistor lines with undeformed complex structure $\bar\partial$.

Classically, this action has a cohomological gauge symmetry associated to gauge transformations of $\tilde h$.
    \begin{equation}
        \tilde h \rightarrow \tilde h + \bar\nabla \tilde \phi, \quad \tilde \phi \in \Omega^{0}(\mathbb{PT}^{\mathcal{O}},\mathcal{O}(-6))\,.
    \end{equation}
Classically, for infinitesimal gauge transformations $\tilde \eta$, there is a cohomological gauge symmetry associated with gauge transformations of $h$
    \begin{equation}
        h \rightarrow h + \bar\nabla \tilde \eta, \quad \tilde h \rightarrow \tilde h - \{\tilde h, \tilde \eta\}\,.
    \end{equation}
It can be shown \cite{Bittleston:2022nfr} that the twistor action is classically equivalent to the CDJ action for self-dual gravity \cite{Capovilla:1991kx} on real Euclidean spacetime
    \begin{equation}
        \int_{\mathbb{R}^4} \psi^{\alpha \beta \gamma \delta} \, \d q_{\alpha \beta} \wedge \d q_{\gamma \delta}\,.
    \end{equation}
In which $q_{\alpha \beta}$ is a spacetime 1-form symmetric on spinor indices, and $\psi^{\alpha \beta \gamma \delta}$ is a 0-form totally symmetric on spinor indices. The direct and indirect Penrose transforms that relate the on-shell twistor fields to the on-shell spacetime fields is
    \begin{equation}
        \int_{L_x} \D \lambda \tilde h \lambda^{\alpha}\lambda^{\beta}\lambda^{\gamma}\lambda^{\delta} = \psi^{\alpha \beta \gamma \delta}(x), \quad \quad \int_{L_x} \frac{\D \lambda \D \hat \lambda}{\langle \lambda \hat \lambda \rangle^2} \frac{\hat \lambda^{\alpha}\hat \lambda^{\beta}}{\langle \lambda \hat \lambda \rangle^2}h = q^{\alpha \beta}\,,
    \end{equation}
where $L_x$ refers to the $\mathbb{CP}^1\in\mathbb{PT}^\cO$ that corresponds to a point $x$ on spacetime through the incidence relation \eqref{incidence_relation}. On spacetime, $\psi^{\alpha \beta \gamma \delta}$ is to be interpreted as the linearised undotted Weyl spinor, while $q^{\alpha \beta}$ is the 1-form encoding the connection on the undotted spin bundle. The metric can be extracted from the triplet of 2-forms $\d q^{\alpha \beta}$ by a magical formula due to Urbantke \cite{Urbantke_1984}:
    \begin{equation}
        \sqrt{g} g_{\mu \nu}(q) = \epsilon^{\sigma \rho \tau\theta} (\d q^{\alpha \beta})_{\mu \sigma}(\d q^{\beta}_{\gamma})_{\rho \tau}(\d q^{\gamma \alpha})_{\theta \nu}\,.
    \end{equation}
It has been observed \cite{Bittleston:2022nfr,costello2021quantizing} that the twistor action has an anomaly, being the standard chiral anomaly in $4k+2$ dimensions. The anomaly is associated with the failure of gauge invariance of the 1-loop 4 point all $h$ process. An anomaly cancellation mechanism reminiscent of the Green-Schwarz mechanism was proposed in section 5.2 of reference \cite{Bittleston:2022nfr}. A new matter field $\rho$ is coupled with the existing fields with a coupling coefficient $\kappa = \mathcal{O}(\sqrt{\hbar})$:
    \begin{equation}
        S_{\text{counter}}= \int_{\mathbb{PT}^\cO} \rho \partial \bar\nabla \rho + \kappa \int_{\mathbb{PT}^\cO} \partial \rho \{\{h,\partial h\}\}, \quad \rho \in \Omega^{1,1}(\mathbb{PT})\,.
    \end{equation}
where $\{\{\cdot,\cdot\}\}$ symbolises a nested Poisson bracket as in \eqref{poisson_bracket}. At 4 points with 4 external $h$, the tree level exchange of the $\rho$ field (being an $\mathcal{O}(\hbar)$ process) precisely cancels the 1-loop 4 point all $h$ process of the original theory, provided that the coupling constant $\kappa$ is chosen to satisfy:
    \begin{equation}
        \kappa^2 = \frac{1}{5!}\left(\frac{\im}{2 \pi}\right)^2\,.
    \end{equation}
The resulting theory $S_{\text{SD}}+S_{\text{counter}}$
is non-anomalous, and some computation shows that it is in fact tree-exact. We shall use this as the definition of our twistor action for self-dual Einstein gravity:
\begin{equation}
    S_{\text{SDG}} =\int_{\mathbb{PT}^\cO} \D^3Z \wedge \tilde h \wedge \left(\bar \partial h + \frac{1}{2}\{h,h\}\right)+ \int_{\mathbb{PT}^\cO} \rho \partial \bar\nabla \rho + \kappa \int_{\mathbb{PT}^\cO} \partial \rho \{\{h,\partial h\}\}\,.
\end{equation}
Observables with no external $\rho$ fields computed in this theory agree precisely with the tree level results from the unmodified self-dual gravity action. The new terms in the action that depend on the new matter field $\rho$ descend to the following terms on spacetime:
    \begin{equation}
        \int_{\mathbb{R}^4} \rho \square_q^2 \rho + \kappa \rho R^\mu_{\nu} \wedge R^{\nu}_{\mu}\,,
    \end{equation}
in which $R^\mu_\nu$ is the Riemann tensor written as a 2 form taking values in endomorphisms of the tangent bundle, and $\square_q$ is the Laplacian for the spacetime defined by $g_{\mu \nu}(q^{\alpha\beta})$.

\section{$\mathbb{C}^*$ scaling reduction and the "splitting" formalism}\label{sec:review_reduction}

\subsection{The "splitting" formalism}\label{subsec:split_idea}
In the context of celestial holography on $\mathcal{M^{O}}:=\mathcal{M}\setminus \{x^{\alpha\dal}=0\}$, the on-shell states we wish to consider are not massless plane waves (as suitable for a study of scattering amplitudes) \cite{Pasterski:2016qvg,Pasterski:2017kqt}
\begin{equation}
    \Phi_{k_{\mu}}(x) = e^{\im k \cdot x}, \quad k^2=0 \,,  
\end{equation}
but rather their Mellin transform, known as conformal primary basis states. 
\begin{equation}\label{eq_cpb_scalar}
    \Phi'_{k_{\mu},\Delta}(x) = \frac{(-\im)^\Delta\,\Gamma(\Delta)}{(k \cdot x)^{\Delta}}, \quad k^2=0,\, \Delta \in \mathbb{C}\,.
\end{equation}
One crucial difference with the massless plane waves is that the massless conformal primary basis states are arbitrarily singular on the locus $\{k\cdot x=0\}$ for generic $\Delta$, and generically do not admit a Fourier decomposition.


Conventionally, the Mellin transform to go from a scattering amplitude to a celestial correlator is taken on on-shell states at the end of a standard scattering amplitude calculation on $\mathcal{M^O}$. In this work as in our previous work \cite{Bu:2023cef, Bu:2023vjt}, we advocate for interpreting the Mellin transform as a decomposition of 4d quantum fields into their scaling modes, integrating out the scaling direction, and working directly on $\mathcal{M^{O}}/\{x \sim s x, s \in \mathbb{R}^+\}$.

One obstacle to this interpretation is that we are forced to enlarge the space of allowed field configurations in our quantum field theory to include not just Fourier decomposable field configurations on $\mathcal{M^O}$, but also field configurations that have arbitrarily singular behaviour on the locus $\{k \cdot x = 0\}$, for some real Lorentzian null vector $k^{\mu}$. 
\begin{equation}
    \underbrace{\int \D \Phi_{k_{\mu}} \, \e^{S[\Phi_{k_{\mu}}]}}_{\text{Fourier states}} \xrightarrow{\text{Mellin mode decomposition}} \int \D\Phi_{k_{\mu}}\,\underbrace{\e^{S[\Phi_{\Delta}]}}_{\text{boost states}} \,.
\end{equation}
Note that the original integration domain is over the space of Fourier decomposable states $\Phi_k$, and this procedure therefore cannot be done. We would like to relate the original theory to a new theory in which the integration domain runs over the boost eigenstates, allowed to have arbitrarily singular behaviour on the locus of $k\cdot x=0$, such that the path integral becomes 
\begin{equation}
    \underbrace{\int \D \Phi'_{\Delta} \, \e^{S[\Phi']_{\Delta}}}_{\text{boost states}}\,.
\end{equation}
Summing over a shifted set of allowed eigenstates $\Phi'_{\Delta}$. To elaborate on this point, consider the usual Fourier decomposition of a free massless scalar action:
\begin{equation}
    \int_{\mathbb{R}^4} \D \Phi(x) \, \exp \left({-\int_{\mathcal{M^O}} \d^4 x \,\Phi \square \Phi}\right) = \int_{\mathbb{R}^4} \D \Phi(k^{\mu}) \, \exp \left({-\int \d^4 k \, k^{\mu}k_{\mu} \Phi(k^{\mu}) \Phi(-k^{\mu})}\right)\,.
\end{equation}
Implicit in the decomposition is the assumption that the original path integral over field configurations was the space of functions subject to Fourier decomposition (or for non conformal theories, that the excluded field configurations are washed out by RG). The usual momentum eigenstates clearly live in this space of field configurations. On the other hand, for generic $\Delta$ the massless conformal primary basis states \eqref{eq_cpb_scalar} clearly do not.

We would like to include such field configurations in the space of field configurations to sum over. However, note that it is not possible to characterise functions that have singular behaviour only on the locus $\{k \cdot x = 0\}$ purely from a spacetime condition that only knows about $x^{\mu} \in \mathcal{M^O}$ in a Lorentz invariant way. 
A natural way of describing these singular configurations is to add coordinates to the target space that capture the data in the choice of null $k^{\alpha\dal}$. Since the defining equation for the singular locus is scale invariant, the minimal extra data required is the $S^2 \cong \mathbb{CP}^1_{\lambda_\alpha,\bar\lambda_{\dal}}$, in which $\lambda^\alpha \bar \lambda^{\dal} \propto k^{\alpha\dal}$ define a real null Lorentzian momentum up to scale:
\begin{equation}
    (x^{\alpha \dot \alpha},\lambda_{\beta}) \in \mathcal{M^O} \times \mathbb{CP}^1\,.
\end{equation}
We have used the following definition for Lorentzian conjugations:
\begin{equation}
    \bar \lambda^{\dot \alpha} = \begin{pmatrix}
        \bar \lambda^0 \\ \bar \lambda^1
    \end{pmatrix} ,\quad \bar\mu_{\alpha} =\begin{pmatrix}
        \bar \mu_{\Dot{0}} \\ \bar \mu_{\Dot{1}}
    \end{pmatrix}\,.
\end{equation}
The singular locus can now be written as
\begin{equation}
    \{x^{\alpha \dal}\lambda_{\alpha}\bar \lambda_{\dal}=0\} \subset \mathcal{M^O} \times \mathbb{CP}^1\,.
\end{equation}
This is why we are forced to consider twistor space when we want to describe massless conformal primary basis states and include them into the set of field configurations we wish to integrate over. 

After motivating the use of twistor space from the perspective of Lorentzian Minkowski space, we aim 
to study self-dual gravity in 4d, which is only nontrivial in Euclidean or split signature. To this end, we shall 
set up the formalism to compute in twistor space, $\mathbb{PT} = \mathbb{R}^4 \times \mathbb{CP}^1$. The $\mathbb{CP}^1$ fibre can still be interpreted as encoding real Lorentzian null momenta (up to scale), a special locus in the complexified cotangent bundle over $\mathbb{R}^4$.
In this setting, $x^{\alpha\dal}\lambda_{\alpha}\bar\lambda_{\dal}=0$ is a complex condition setting both $x^{\alpha\dal}\lambda_{\alpha}\bar\lambda_{\dal}$ and its complex conjugate $\hat x^{\alpha\dal}\hat \lambda_{\alpha}\hat{\bar\lambda}_{\dal} = x^{\alpha\dal}\hat \lambda_{\alpha}\hat{\bar\lambda}_{\dal}$ to $0$.  The results of any computation in this signature can be complexified and the real Lorentzian sector can be recovered by imposing appropriate reality conditions. With this in mind, we start by introducing the formalism to include the extra singular configurations that naturally appear in the context of boost eigenstates.

The procedure of going from integrating over Fourier decomposable $\Phi(x)$ to $\Phi'(x,\lambda,\bar\lambda)$ that is permitted to be singular on $\{x^{\alpha \dal}\lambda_{\alpha}\bar \lambda_{\dal}=0\}$ is known as "splitting", where we take a theory on $(\mathbb{R}^4\setminus\{0\})$, promote the theory to one living on $\mathbb{PT}^\cO = (\mathbb{R}^4\setminus\{0\})\times\mathbb{CP}^1$ and split the integral over $\D \bPhi$ into integrals over $\D \bPhi'\D \Lambda$, where $\D \bPhi'$ is over field configurations permitted to be singular and $\Lambda$ is a new auxiliary field that encodes the rescaled on-shell boundary value of $\bPhi'$ on the singular locus. This is best illustrated with an example. Again, consider the massless free scalar theory on spacetime:
\begin{equation}
    \int \D \Phi \, \exp \left(\int_{(\mathbb{R}^4\setminus\{0\})} \d^4 x \Phi \square \Phi\right)\,.
\end{equation}
The first step is to uplift the theory to our twistor space $\mathbb{PT}^\cO$, on which we have the extra coordinates with which to describe the locus $ \{x^{\alpha \dal}\lambda_{\alpha}\bar \lambda_{\dal} = 0\}$:
\begin{equation}
    \int \D \boldsymbol{\Phi} \, \exp \left({\int_{\mathbb{PT}^\cO} \D^3Z \, \bPhi \wedge \bar \partial \bPhi}\right)\,,
\end{equation}
in which $\bPhi \in \Omega^{0,1}(\mathbb{PT}^{\mathcal{O}}, \mathcal{O}(-2))$. The equivalence of this theory to the free scalar theory is explained in appendix \ref{appendix_A}. Crucially, this theory has a gauge symmetry that arises from the nilpotence of the $\bar \partial$ operator:
\begin{equation}
    \bPhi \rightarrow \bPhi + \bar \partial f, \quad f \in \Omega^{0,0}(\mathbb{PT}^\mathcal{O}, \mathcal{O}(-2))\,.
\end{equation}
As a first step towards "splitting" the theory, consider a gauge transformation that is singular on $\{[\mu \bar \lambda] = 0\}$ in a simple way: having a simple pole. Consider the following gauge transformation with an arbitrary function $\Lambda$ (Fourier decomposable on $\mathbb{R}^4 \setminus\{0\}$ for all values of $\lambda^\alpha$):
\begin{align}
    &\bPhi \rightarrow \bPhi + \bar \partial \left(\frac{\Lambda}{[\mu \bar \lambda]}\right) = \bPhi +  \frac{\bar \partial \Lambda}{[\mu \bar \lambda]} + \Lambda \bar \partial_{\lambda}\left(\frac{1}{[\mu \bar \lambda]}\right) + \Lambda \bar \delta([\mu \bar \lambda]) =: \bPhi' + \Lambda \bar \delta([\mu \bar \lambda])\nonumber
    \\
    &\bPhi' := \bPhi +  \frac{\bar \partial \Lambda}{[\mu \bar \lambda]} + \Lambda \bar \partial_{\lambda}\left(\frac{1}{[\mu \bar \lambda]}\right)\,,
\end{align}
where we defined $\bPhi'$ as $\bPhi$ plus the $\Lambda$-dependent simple and double pole. Although $\bPhi$ was Fourier decomposable in the spacetime factor, $\bPhi'$ by construction has at least a single and double pole on the singular locus $\{[\mu \bar \lambda]=0\}$ for generic $\Lambda$. This brings us to the core idea of "splitting":
\begin{enumerate}
    \item Promote $\Lambda$ to a quantum field, i.e. do an additional path integral to sum over $\bPhi$ field configurations which are gauge equivalent with respect to a gauge transformation with a simple pole at $[\mu\bar\lambda]=0$.
    \item Perform a change of integration variable from $\bPhi$ (Field configurations Fourier decomposable in $x$) to $\bPhi'$ (Field configurations permitted to be have a simple and double pole at $\{[\mu \bar \lambda]=0\}$).
\end{enumerate}
The "split" theory is therefore:
\begin{align}
    &\int \D \boldsymbol{\Phi} \, \exp \left({\int_{\mathbb{PT}^\cO} \D^3 Z \, \bPhi \wedge \bar \partial \bPhi}\right) \longrightarrow\nonumber\\
    &\int \D \bPhi' \D \Lambda \, \exp{\left(\int_{\mathbb{PT}^\cO} \D^3 Z \, (\bPhi' + \Lambda \bar \delta([\mu \bar \lambda]) \wedge \bar \partial (\bPhi' + \Lambda \bar \delta([\mu \bar \lambda]) \right)} \nonumber
    \\
     = &\int \D \bPhi' \D \Lambda \, \exp{\left(\int_{\mathbb{PT}^\cO} \D^3 Z \, \bPhi' \wedge \bar \partial \bPhi' + 2\int_{\mathbb{PT}^\cO} \D^3Z \wedge \bar \delta([\mu \bar \lambda])\Lambda \wedge \bar \partial \bPhi' \right)}\,, \label{eq:split_scalar}
\end{align}
where the term quadratic in $\Lambda$ vanishes because of skew symmetry of the wedge product. The reader unfamiliar with this notation is directed to appendix \ref{appendix_A}. Note the action functional now splits into one part living on $\mathbb{PT}$ summing over field configurations including singular ones on the locus $[\mu\bar\lambda]=0$, while a second part localises to that locus with a free functional degree of freedom $\Lambda$ summed over. 

In step 1, since $\Lambda$ arose as a gauge transformation, (for a non-anomalous theory such as this free theory) the theory is unchanged by the new path integral over $\D \Lambda$. This is because $\Lambda$ can (by definition) be decoupled by a gauge transformation of $\bPhi$, so that the $\int \D \Lambda$ only provides a volume factor of the gauge group which is cancelled by the same factor from the partition function. In step 2, since this field redefinition is linear in $\bPhi$, the path integral Jacobian to go from $\D \bPhi \D\Lambda \rightarrow \D \bPhi'\D\Lambda$ is trivial. Crucially, despite appearances, the space of field configurations of $\bPhi'$ to sum over does not depend on $\Lambda$, which is a special feature of the free theory and the BF theory we will consider for self-dual gravity. This is because the path integral evaluations reduce to functional determinants evaluated on an on-shell background, and the on-shell conditions restrict $\bPhi'$ to precisely the locus in field configuration space whose singularity structure is controlled by $\Lambda$:
\begin{equation}
    \bar \partial \left(\bPhi' + \Lambda \bar \delta([\mu \bar \lambda])\right) = 0 \implies \bPhi' = (\text{hol. (0,1) forms}) + \frac{\bar \partial \Lambda}{[\mu \bar \lambda]} + \Lambda \bar \partial_{\lambda}\left(\frac{1}{[\mu \bar \lambda]}\right)\,.
\end{equation}
The particular solution can be constructed by inspection, and the complementary solution is simply the space of $\bar \partial$-closed $(0,1)$-forms of weight $-2$ in $\lambda^{\alpha}$. Therefore there is no difference in any computation if we allow the path integral over $\bPhi'$ to be over all field configurations with arbitrary simple and double poles, as the path integral will in any case holomorphically localise to the locus in field configuration space where the singularities of $\bPhi'$ are controlled by $\Lambda$ in the above way.

This theory enjoys a new shift symmetry:
\begin{equation}
    \bPhi' \rightarrow \bPhi' + \bar \delta([\mu \bar \lambda])f, \quad \Lambda \rightarrow \Lambda - f\,.
\end{equation}
Observables computed using operator insertions that have this shift symmetry coincide precisely with observables computed using the original $\bPhi$. This is because operator insertions that respect this shift symmetry are built out of $(\bPhi' + \Lambda \bar \delta([\mu \bar \lambda])$, and the "split" action has the same functional dependence on $(\bPhi' + \Lambda \bar \delta([\mu \bar \lambda])$ as the original action did on $\bPhi$.

The crucial new behaviour in the "split" theory that cannot be easily reproduced in the original theory arises from operator insertions that do not enjoy this shift symmetry, and therefore are not immediately recognisable as objects in the original theory\footnote{For example, a bare $\Lambda$ or a bare $\bPhi'$.}. Analysis of correlators of such fields is the key behind the results in this work.

It is worth spending some time now to properly understand the physical interpretation of what we have done in going to the "split" theory, as well as the physical role of the quantum field $\Lambda$.

The motivation to go to twistor space and to "split" the theory comes from wanting to include the massless conformal primary basis states into the space of field configurations we wish to integrate over, so that we can perform the decomposition into scaling modes at the level of the Lagrangian rather than after having evaluated a 4d theory correlator or amplitude. The locus $\{[\mu \bar \lambda]=0\}$ has been picked out to be special by the form of the massless conformal primary basis states. These $\Lambda$ were cohomological gauge transformations that localise to the singular locus. Being gauge transformations, some parts of $\Lambda$ are trivial: small gauge transformations that need to be gauge fixed away. However, some parts of $\Lambda$ are genuinely physical field configurations, being gauge transformations that do not vanish in the large radius limit in the $\mathbb{R}^4\setminus\{0\}$ factor. By "splitting" the theory, we have picked out such gauge transformations and pulled them out of the path integral $\D \bPhi$, placing them on even footing with the original bulk field. Furthermore, we have enlarged the space of field configurations to $\D\bPhi'$ that can be singular on the locus $\{[\mu \bar \lambda]=0\}$, so that the massless conformal primary states \eqref{eq_cpb_scalar} are now in the space of allowed field configurations.

By performing these manipulations, we have now set the theory up in a way that is well adapted to the study of conformal primary states and their allowed large gauge transformation on spacetime. From the above arguments, in the sector where we compute with observables that respect the shift symmetry, the "split" theory is identical to the original theory. However, observables in the "split" theory that do not respect the shift symmetry let us directly study, for example, large gauge transformations that live on the celestial sphere. Furthermore, on-shell massless conformal primary basis states \eqref{eq_cpb_scalar} can now live in the field configuration space, and do not have to be defined by a limiting procedure.

In practice, we want to consider boost eigenstates with $\Delta\neq 1$, which suggests that we should allow for arbitrarily singular on-shell field configurations. The twistor wavefunctions that correspond to the on-shell massless conformal primary wavefunctions can have dependence on $[\mu \bar \lambda]^{k}$ for generic $k \in \mathbb{C}$. We therefore consider gauge transformations $\Lambda_{k}$ with arbitrary $k$ and sum/integrate over all values of $k$ that we wish to include in the "split" theory:
\begin{align}
    &\bPhi \rightarrow \bPhi + \bar \partial \left(\frac{\Lambda_{k}}{[\mu \bar \lambda]^{k}}\right) = \bPhi +  \frac{\bar \partial \Lambda_{k}}{[\mu \bar \lambda]^{k}} + \Lambda_{k} \bar \partial_{\lambda}\left(\frac{1}{[\mu \bar \lambda]^{k}}\right) + \Lambda_{k} \bar \partial_{\mu}\left(\frac{1}{[\mu \bar \lambda]^{k}}\right)\nonumber
    \\
    &\bPhi' := \bPhi +  \frac{\bar \partial \Lambda_{k}}{[\mu \bar \lambda]^{k}} + \Lambda_{k} \bar \partial_{\lambda}\left(\frac{1}{[\mu \bar \lambda]^{k}}\right)\,. \label{3.14}
\end{align}
In complete analogy with the previous simple case, we also change the integration measure to include the modified field configurations and the new quantum field $\Lambda$: $\D \bPhi \rightarrow \D \bPhi' \D \Lambda_{k}$. Note that only for $k \in \mathbb{Z}^+$ do we get localisation to formal neighborhoods of the locus $\{[\mu \bar \lambda] = 0\}$.

\subsection{$\mathbb{C}^*$-scaling reduction}
After introducing the boost eigenstates of interest by performing almost trivial gauge transformations on $\mathbb{PT}^\cO$ that splits the theory on the level of the action functional \eqref{eq:split_scalar}, we would like to isolate the dynamics of the free functional degree of freedom $\Lambda$ on the locus $[\mu\bar\lambda]=0$. We choose to do a further $\mathbb{C}^*$ scaling reduction by giving independent projective scaling to $\mu^{\dal}$, which takes our "split" theory down
\begin{equation}
    \mathbb{CP}^1\times\mathbb{CP}^1\cong\mathbb{MT}\oplus\mathbb{CP}^1_{[\mu\bar\lambda]=0}=\frac{\mathbb{PT}^{\cO}}{\{\mu^{\dal}\sim s \mu^{\dal}, s \in \mathbb{C}^*\}}\,.
\end{equation}
As discussed in our previous papers \cite{Bu:2023cef,Bu:2023vjt}, $\mathbb{MT}\cong\mathbb{CP}^1\times\mathbb{CP}^1\setminus\mathbb{CP}^1_{[\mu\bar\lambda]=0}$ is the space of oriented geodesics of EAdS$_3=\mathbb{H}_3$ or oriented timelike geodesics on LdS$_3$ slices of Minkowski space, while $\mathbb{CP}^1_{[\mu\bar\lambda]=0}$ is the space of degenerate geodesics (that start and end at the same point) on their common boundary $S^2$ located at $u=0$. An discussion of how this relates to the Lorentzian signature $\mathcal{M}^{\cO}\times\mathbb{CP}^1$ can be found in appendix \ref{appendix:Lorentzian_story}. The reason we do not work in Lorentzian signature throughout is that unfortunately $\mathcal{M}^{\cO}\times\mathbb{CP}^1$ is not an interesting setting for us, as the self-duality condition implies (conformal) flatness. Here we focus on the $(\mathbb{R}^4\setminus\{0\}) \times \mathbb{CP}^1 = \mathbb{PT}^\cO$ where practical calculations can be carried out. Another benefit to this choice of signature is that $(\mathbb{R}^4\setminus\{0\}) \times \mathbb{CP}^1$ is a complex 3-fold, so working and notation is cleaner. 

\begin{figure}
    \centering
    \adjustbox{scale=1.3,center}{
\begin{tikzcd}[row sep=huge]
 & \mathbb{PT}^{\cO} \ar[ld,"u=0"] \ar[rd,"u\neq 0"] \\
 \{q,\bar q,\lambda,\bar\lambda\} \ar[d,"\mathcal{O}_q(2)"]  & & \{q,\bar q,u,\bar u,\lambda,\bar\lambda\} \ar[d,"\mathcal{O}(1\text{,}-1)"]\\
 \mathbb{CP}^1_{[\mu\bar\lambda]=0} & & \mathbb{MT}\cong\mathbb{CP}^1_{\nu}\times\mathbb{CP}^1_{\lambda}\setminus\mathbb{CP}^1_{[\nu\bar\lambda]=0} 
\end{tikzcd}}
    \caption{A summary of the geometry involved in $\mathbb{C}^*$-scaling reduction of the split theory.}
    \label{fig:reduction}
\end{figure}
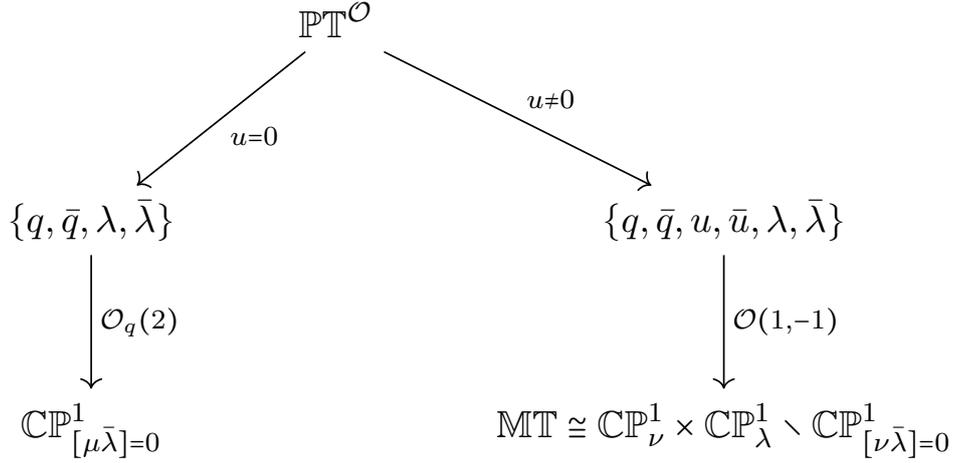

In preparation for a $\mathbb{C}^*$ reduction, consider the following coordinates for $\mathbb{PT}^\cO$ as summarised in Figure \ref{fig:reduction}, 
\begin{align}
    &\{u, \bar u, q, \bar q, \lambda^{\alpha}\} \in \mathbb{PT}^\cO \nonumber
    \\
    & u = \frac{[\mu\bar\lambda]}{\la\lambda\hat{\lambda}\ra} \in \mathcal{O}, \quad q=[\mu\hat{\bar\lambda}] \in \mathcal{O}(2)\,,
\end{align}
in which $u, q$ are not permitted to simultaneously vanish, as $\mu^{\dal} \neq 0$. The locus $\{u=0\} \subset \mathbb{PT}^\cO$ is therefore the total space of an $\mathcal{O}(2)$ bundle over $\mathbb{CP}^1$, with fibre coordinate $q\neq 0 $ being a $\mathbb{C}^*$ valued section of $\mathcal{O}(2)$. 
\begin{equation}
    \mathcal{O}_{q\neq0}(2) \rightarrow \{u=0\} \rightarrow \mathbb{CP}_{\lambda}^1\,.
\end{equation}
Along similar lines, the locus $\{u \neq 0\}$ can be interpreted as the total space of a $\mathcal{O}(1_\lambda,-1_\nu)$ bundle over $\mathbb{CP}^1_{\nu} \times \mathbb{CP}^1_{\lambda}\setminus \mathbb{CP}^1_{[\nu \bar \lambda]=0}$.
Having removed the locus $[\nu \bar \lambda]=0$ as well, we can write the change of coordinates from $\nu^{\dal}$ to $\mu^{\dal}$ as
\begin{equation}
    \frac{[\mu \bar \lambda]}{[\nu \bar \lambda]} v^{\dal} = \mu^{\dal}\,.
\end{equation}
We can redefine $u := \frac{[\mu \bar \lambda]}{[\nu \bar \lambda]} $ which manifestly has weight $\mathcal{O}(1_{\lambda}, -1_{\nu})$ and is nonzero. We have the trivial fibration:
\begin{equation}
    \mathcal{O}_{u \neq 0}(1_\lambda,-1_\nu)\rightarrow \{u \neq 0 \} \rightarrow \mathbb{CP}^1_{\nu}\times \mathbb{CP}^1_{\lambda} \setminus \mathbb{CP}^1_{[\nu\bar\lambda]=0}\,.
\end{equation}
Note that the removed diagonal $\mathbb{CP}^1_{[\nu\bar\lambda]=0}$ is (up to a coordinate reparameterisation), the same $\mathbb{CP}^1_{[\mu\bar\lambda]=0}$  on the left part of the diagram.
In each patch, we have isolated the coordinate that scales under
\begin{equation}
    \mu^{\dal} \rightarrow c\mu^{\dal}, \quad c \in \mathbb{C}^*\,,
\end{equation}
and written it as a $\mathbb{C}^*$ valued fibre coordinate over a real codimension 2 base space. We will want to decompose our fields into basis functions that encode the dependency on the fibre coordinate (and therefore all the dependence on the scale of $\mu^{\dot \alpha}$), and then integrate out the fibre coordinate. For instance, consider the integrand $I(Z^A, \bar Z^A)$ which is of the form of products of fields, we decompose each field in orthogonal basis function:
\begin{equation}\label{eq_B_basis}
    B_{s,m}(u):=(u \bar u)^{\im s} \left(\frac{u}{\bar u}\right)^{m/2}, \quad s \in \mathbb{R}, m \in \mathbb{Z}\,.
\end{equation}
In local coordinates on $\mathbb{C}^*$ coordinatised by $(\rho = \log r, \theta)$ in which $(r, \theta) \in \mathbb{R}^+ \times S^1 = \mathbb{C}^*$, the basis functions shown are the canonical orthogonal Fourier transform/series basis functions:
\begin{equation}
    B_{s,m}|_{\text{local}} = e^{\im s \rho} e^{\im m \theta}, \quad s \in \mathbb{R}, m \in \mathbb{Z}\,.
\end{equation}
Using this, the integrand can be written as
\begin{equation}
\int_{\mathbb{PT}^\cO} \frac{\D^3 Z \D^3 \bar Z}{\la \lambda \hat \lambda \ra^4}\,  I(Z^A) =
\begin{cases}
        \int_{\mathbb{PT}^\cO} \frac{(\D \nu \D \hat \nu) \D \lambda \D \hat \lambda(u \bar u \d u \d \bar u)}{\la \lambda \hat \lambda \ra^4}  \left(\int_{\mathbb{R}} \d s \sum_{m}I_{(s,m)}(\nu^{\dal},\lambda_{\alpha}) B_{s,m}(u)\right)\,, & u\neq 0\,;\\
        \int_{\mathbb{PT}^\cO} \frac{\d q \d \bar q \D \lambda \D \hat \lambda}{\la \lambda \hat \lambda \ra^4}\, \left(\int_{\mathbb{R}} \d s \sum_{m} I_{(s,m)}(\lambda)B_{s,m}(q)\right)\,, & u=0 \,,
\end{cases}
\end{equation}
where we have suppressed the sum over the basis function for individual component field in $I$. Using orthogonality of the basis functions, the fibre can be integrated out, setting the total $s$ and total $m$ to 0 for the products of fields in $I$. We collectively label the mode numbers as $\Delta=m+\im s\in\mathbb{Z}+\im\mathbb{R}$ in the following sections. $\Delta$ labels the $\mathbb{C}^*$-scaling in $\mu^{\dal}$, the real part of which is the usual Mellin mode in the scaling in $r\sim |x|$ on spacetime. Note that our $\Delta\in \mathbb{Z}+\im\mathbb{R}$ contains both the usual principle series $1+\im\mathbb{R}$ and the integer $\mathbb{Z}$ soft/Goldstone modes. 

In the subsequent discussion, we perform this procedure for self-dual gravity on $\mathbb{PT}^\cO$. To recap, the roadmap is:
\begin{enumerate}
    \item Begin with the twistor action \eqref{hth_twistor_action} on $\mathbb{PT}^\cO$ and perform gauge transformations \eqref{3.14} to induce arbitrarily singular behaviour on the defect locus $[\mu\bar\lambda]=0$. 
    \item With proper field redefinitions and localisation, formally split the theory into two parts.
    \item In order to extract dynamics of the pure gauge function, perform $\mathbb{C}^*$-scaling reduction to obtain action functionals on $\mathbb{MT}$ and $\mathbb{CP}^1_{[\mu\bar\lambda]=0}$, with all the fields labelled by mode numbers $\mathbb{Z}+\im\mathbb{R}$. 
    \item Study the 2d theory on $\mathbb{CP}^1_{[\mu\bar\lambda]=0}$ and its marginal deformations.
\end{enumerate}
We shall follow this roadmap in the remainder of the paper to analyse the self-dual action on $\mathbb{PT}^\cO$ in the context of celestial holography.

\section{A case study: self-dual gravity}\label{sec:reduction_2d}
\subsection{Large pure diffeomorphism}\label{subsec:edge_modes}
We consider linearised gauge transformations 
in solutions of vacuum Einstein equation in 4d Minkowski space of the form 
\begin{equation}
    \delta h_{\mu\nu} = \partial_{(\mu}H_{\nu)}\,,
\end{equation}
in which $H_{\nu}$ is a function encoding a pure diffeomorphism, which our first instinct would be to quotient out when considering the space of metric configurations. One particularly interesting scenario is when $H_{\nu}$ scales like $\cO(r^0)$ in a large $r\propto |x|$ limit. Such pure diffeomorphism generators generate gauge-identical field configurations in the bulk but do not vanish asymptotically in a large $r$ expansion, which means they 
act nontrivially on boundary conditions or boundary operator insertions on the conformal boundary $\scri$. One particular example of such generators were considered in \cite{Donnay:2018neh}: 
\begin{equation}\label{large_gauge_profile_1}
    \delta h_{\mu\nu}=\partial_{(\mu}\partial_a^2\left[q_{\nu)}\,\log(-q\cdot x)\right] = \partial_{(\mu}\left[\frac{2\partial_a q_{\nu)}\partial_a q\cdot x}{q\cdot x}-\frac{ q_{\nu)}(\partial_a q\cdot x)^2}{(q\cdot x)^2}\right]=:\partial_{(\mu}\xi_{\nu)}\,,
\end{equation}
with $\partial_a q_{\mu}=\veps_{\mu}=\kappa_{\alpha}\hat{\tilde\kappa}_{\dal}$, a null polarization vector. Notice that $\xi_{\nu}$ is homogeneous of weight $0$ in $x^{\alpha \dot \alpha}$. Hence if one were to allow the existence of such large pure diffeomorphism profiles, we are forced to include bulk field configurations which falls off to different boundary values in a sum over field configurations. Although the notion of summing over field configurations is subtle in GR due to the well-known renormalizability issues with the integration measure in path integrals, we avoid this problem using the mechanism of twistor theory to consider an equivalent holomorphic Poisson BF theory which descends to SDGR in 4d. The path integral for the holomorphic Poisson BF theory on $\mathbb{PT}^\cO$ is non-anomalous, and it would be interesting to study how the inclusion of the extra field configurations generated by boundary large diffeomorphism profiles on spacetime affect the theory. To this end, we first consider the uplift of the large pure diffeomorphism profile \eqref{large_gauge_profile_1} to $\mathbb{PT}^\cO$. 

We consider the Sparling transform of the following profile on $\mathbb{PT}^\cO$:
\begin{equation}\label{PT_lgp}
  \bar\partial\left(\frac{\la\lambda\hat{\lambda}\ra[\mu\hat{\bar\lambda}]^2}{[\mu\bar\lambda]}\tilde\eta\right)\,,
\end{equation}
where $\tilde\eta\in\Omega^0\otimes\cO(-2)$ is required to be independent of $\hat \mu^{\dal}$ and has projective weight $-2$. (In practice we will usually use $\tilde \eta(\lambda, \bar \lambda)$ that only depends on $\mathbb{CP}^1_{\lambda}$, and is wholly independent of $\mu^{\dal}, \hat \mu^{\dal}$). Its spacetime counterpart can be computed with the following integral transform for the linearised metric perturbation:
\begin{equation}
    \left.\int_{\mathbb{CP}^1} \D\lambda \frac{\iota_{\alpha}\xi_{\beta}}{\la\lambda\iota\ra\la\lambda\xi\ra}\,\frac{\partial^2}{\partial\mu^{\dal}\partial\mu^{\Dot{\beta}}}\,\bar\partial\left(\frac{\la\lambda\hat{\lambda}\ra[\mu\hat{\bar\lambda}]^2}{[\mu\bar\lambda]}\tilde\eta\right)\right\vert_{L_x} \,,
\end{equation}
where $\iota^{\alpha}$ and $\xi^{\beta}$ are a pair of arbitrary reference spinors that encode a gauge choice. By turning one of the $\mu^{\dal}$ derivatives into a spacetime derivative
\begin{equation}
    \frac{\partial}{\partial\mu^{\dal}}=\frac{\hat{\iota}^{\alpha}}{\la\lambda\hat{\iota}\ra}\,\frac{\partial}{\partial x^{\alpha\dal}}\,,
\end{equation}
(this is true when these differential operators act on $\tilde \eta$, as it is independent of $\hat \mu^{\dal}$). The spacetime derivative can be pulled outside the integral. We then integrate by parts on the $\bar\partial$, which can turn into holomorphic delta functions of either $\bar\delta(\la\lambda\iota\ra)$ or $\bar\delta(\la\lambda\xi\ra)$.
\begin{equation}
    \left.\frac{\partial}{\partial x^{\alpha\dal}}\int_{\mathbb{CP}^1}\D\lambda\frac{\la\iota\hat{\iota}\ra\xi_{\beta}}{\la\lambda\hat\iota\ra}\,\frac{\partial}{\partial\mu^{\dbl}}\left(\frac{\la\lambda\hat{\lambda}\ra[\mu\hat{\bar\lambda}]^2}{[\mu\bar\lambda]}\tilde\eta\right)\right\vert_{L_x}\wedge\left(\frac{\bar\delta(\la\lambda\iota\ra)}{\la\lambda\xi\ra}+\frac{\bar\delta(\la\lambda\xi\ra)}{\la\lambda\iota\ra} \right)\,.
\end{equation}
Evaluating the $\mu$-derivative and doing the integral against the holomorphic delta functions gives us the large pure diffeomorphism graviton profile \eqref{large_gauge_profile_1}, with appropriate choices of the two pairs of reference spinors to be chiral part of the external momenta $\iota^{\alpha}=\kappa^{\alpha}$, $\xi^{\alpha}=\hat{\kappa}^{\alpha}$ and if $\tilde \eta$ was independent of $\mu$, the spacetime function that $\tilde \eta$ descends to is a $\kappa$ or $\hat \kappa$ dependent function that is independent of $x$.

Yet another perspective on the $\bar\partial$-exact profile \eqref{PT_lgp} on $\mathbb{PT}^\cO$ one could take is to consider it as a localising holomorphic delta function on the locus of $[\mu\bar\lambda]=0$. This amounts to recognizing this profile as a (linearised) holomorphic gauge transformation for $h\in\Omega^{0,1}\otimes\cO(2)$:
\begin{equation}
    h\to h+  \bar\partial\left(\frac{\la\lambda\hat{\lambda}\ra[\mu\hat{\bar\lambda}]^2}{[\mu\bar\lambda]}\tilde\eta\right)\,.
\end{equation}
This is quite analogous to the discussion we have had on spacetime, rather instead of a large diffeomorphism generator, we have a almost everywhere trivial gauge transformation for the off-shell bulk graviton on $\mathbb{PT}^\cO$ except on the locus of $\mathbb{CP}^1_{[\mu\bar\lambda]=0}$. Analogous to the case in 4d general relativity
\begin{equation}
    \boxed{\frac{g_{\mu\nu}}{\text{diffeo}} + \text{large pure diffeo}} \xRightarrow{\text{uplift}} \boxed{\frac{h}{\bar\partial-\text{exact}} +\bar\delta([\mu\bar\lambda])\tilde\eta}\,.
\end{equation}
If we allow such almost everywhere $\bar\partial$-exact gauge transformations, the sum over field configuration is required to include further configurations which only differ on their values on $\mathbb{CP}^1_{[\mu\bar\lambda]=0}$. Since such field configurations are simply labelled by the function $\tilde\eta$, we promote $\tilde\eta$ to a quantum field labelling field configurations of bulk $h$ which only differ on the defect $\mathbb{CP}^1_{[\mu\bar\lambda]=0}$ by $\tilde\eta$. We shall see that these profiles are allowed to have independent dynamics on their own, which reflects the bulk dynamics in a holographic way.

We are also free to consider other profiles on $\mathbb{PT}^\cO$:
\begin{equation}\label{PT_lgp_k}
    \bar\partial\left(\frac{\la\lambda\hat{\lambda}\ra^k[\mu\hat{\bar\lambda}]^{k+1}}{[\mu\bar\lambda]^k}\tilde\eta_{-2k}\right)\,,\quad k\in\mathbb{Z}_+\,,
\end{equation}
such that the entire profile takes values in $\Omega^{0,1}\otimes\cO(2)$. On spacetime, their counterpart takes the form of a generalised large pure diffeomorphism generator: 
\begin{equation}
     \delta h_{\mu\nu}^{(k)} = \partial_{(\mu}\left[\frac{(k+1)\partial_a q_{\nu)}(\partial_a q\cdot x)^k}{(q\cdot x)^k}-\frac{k\, q_{\nu)}(\partial_a q\cdot x)^{k+1}}{(q\cdot x)^{k+1}}\right]\,,\quad k\in\mathbb{Z}_+\,.
\end{equation}
It is evident that such higher pure diffeomorphism profiles have higher order poles in $u=q\cdot x$. On $\mathbb{PT}^\cO$, evaluating the $\bar\partial$ for \eqref{PT_lgp_k} allows us to view it as a holomorphic delta function $\bar\delta^{k-1}([\mu\bar\lambda])$ localising to $(k-1)$th order formal neighborhood of $\mathbb{CP}^1_{[\mu\bar\lambda]=0}$.
\begin{figure}
     \centering
     \begin{subfigure}[ht]{0.45\textwidth}
         \centering
         \includegraphics[width=\textwidth]{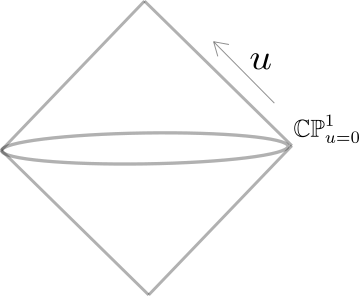}
         \caption{}
         \label{fig:lc1}
     \end{subfigure}
     \hfill
     \begin{subfigure}[ht]{0.5\textwidth}
         \centering
         \includegraphics[width=\textwidth]{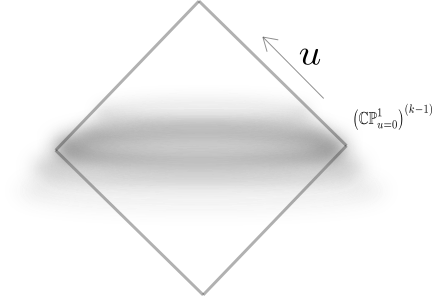}
         \caption{}
         \label{fig:lc2}
     \end{subfigure}
        \caption{Light cone cut at $u=0$ and its $(k-1)$th order formal neighborhood.}
        \label{fig:lcs}
\end{figure}
Consider figure \ref{fig:lcs}, the $\{t \geq 0\}$ Milne region of $\mathcal{M}$. Figure \ref{fig:lc1} shows the light-cone cut at $u=0$, figure \ref{fig:lc2} exhibits the localisation to higher formal neighborhoods of the celestial sphere, which extends in the $u$ direction along $\scri$. When we insert large pure diffeomorphism profiles of all $k\in\mathbb{Z}_+$, the promoted boundary fields are effectively smeared over the null boundary. 

\subsection{CFT and discussion of currents}\label{subsec:2d_CFT_currents}
We wish to perform the splitting procedure as described in \ref{subsec:split_idea} on the CDJM twistor action for sd Gravity. Consider gauge transformations which are singular on the locus $[\mu\bar\lambda]=0$
\begin{equation}
    \tilde h\to \tilde h+\bar\nabla\left(\frac{q^{k+1}\la\lambda\hat{\lambda}\ra^k\,\tilde\phi}{[\mu\bar\lambda]^k}\right) -\left\{\tilde h, \left(\frac{q^{k+1}\la\lambda\hat{\lambda}\ra^k\,\tilde\eta}{[\mu\bar\lambda]^k}\right)\right\} \,,\quad h\to h+\bar\nabla\left(\frac{q^{k+1}\la\lambda\hat{\lambda}\ra^k\,\tilde\eta}{[\mu\bar\lambda]^k}\right)\,.
\end{equation}
The $\bar\partial$ part of the gauge transformation contains a localising part and a non-localising part:
\begin{equation}
    \bar\partial\left(\frac{q^{k+1}\la\lambda\hat{\lambda}\ra^k\,\tilde\phi}{[\mu\bar\lambda]^k}\right) = \bar\delta^{(k-1)}([\mu\bar\lambda])q^{k+1}\la\lambda\hat{\lambda}\ra^k\,\tilde\phi + \frac{\bar\partial\left(q^{k+1}\la\lambda\hat{\lambda}\ra^k\,\tilde\phi\right)}{[\mu\bar\lambda]^k} -k\, \frac{q^{k+2}\la\lambda\hat{\lambda}\ra^k\tilde\phi}{[\mu\bar\lambda]^{k+1}}\,.
\end{equation}
Following the splitting procedure outlined in \ref{subsec:split_idea}, we shift bulk fields to absorb the non-localising part:
\begin{equation}\label{eq:field_redef}
\begin{split}
    &h'=h + \frac{\bar\nabla\left(q^{k+1}\la\lambda\hat{\lambda}\ra^k\,\tilde\eta\right)}{[\mu\bar\lambda]^k} -k\,\left(\left[\frac{\partial h}{\partial\mu}\bar\lambda\right]+\tilde\eta\D\bar\lambda\right) \frac{q^{k+2}\la\lambda\hat{\lambda}\ra^k}{[\mu\bar\lambda]^{k+1}} \\
    &\tilde h'=\tilde h +\frac{\left(\bar\nabla-\{\tilde h,\cdot\}\right)\left(q^{k+1}\la\lambda\hat{\lambda}\ra^k\,\tilde\phi\right)}{[\mu\bar\lambda]^k} - k\,\left(\left[\frac{\partial(h-\tilde h)}{\partial\mu}\bar\lambda\right]+\tilde\phi\D\bar\lambda\right) \frac{q^{k+2}\la\lambda\hat{\lambda}\ra^k\tilde\phi\D\bar\lambda}{[\mu\bar\lambda]^{k+1}}\,.
\end{split}
\end{equation}
Crucially, the edge modes $\tilde \phi,\tilde\eta$ are defined to be independent of $[\mu \bar \lambda]$:
\begin{equation}
    \left[\hat{\bar\lambda} \frac{\partial}{\partial \mu}\right]\tilde \phi = \left[\hat{\bar\lambda} \frac{\partial}{\partial \mu}\right]\tilde\eta=0\,.
\end{equation}
Our twistor action splits into two parts, schematically
\begin{equation}
    S[h,\tilde h] = S_{[\mu\bar\lambda]\neq 0}[h',\tilde h'] + S_{[\mu\bar\lambda]=0}[h',\tilde h',\tilde\phi,\tilde\eta]\,.
\end{equation}
More explicitly, we can write down the full action 
\begin{align}
    S_{\text{SDG}'}&= \int_{\mathbb{PT}^\cO}\D^3Z \wedge \tilde h' \wedge \left(\bar \partial h' + \frac{1}{2}\{h',h'\}\right)+ \underbrace{\rho \partial \bar\nabla \rho + \kappa \partial \rho \{h,\partial h\}}_{S_{\text{ct-3d}}}\nonumber\\
    & + \sum_{k\in\mathbb{Z}_+}\int_{\mathbb{PT}^\cO}\D^3Z\wedge \bar\delta^{(k-1)}([\mu\bar\lambda])q^{k+1}\la\lambda\hat{\lambda}\ra^k\left(\tilde\phi\bar\partial h'+\tilde h'\bar\partial\tilde\eta+\tilde\phi\{h',h'\}+\tilde\eta\{\tilde h',h'\}\right)\,,
\end{align}
where the first term is simply rewriting the self-dual gravity action $S_{\text{SDG}}[h,\tilde h]$ and shifting the bulk fields to the redefined $h',\tilde h'$ \footnote{The notation $\bar \delta^{(n)}(z)$ indicates:
\begin{equation}
 \bar \delta^{(n)}(z) = \frac{\partial^n}{\partial z^n} \bar \delta(z), \text{ satisfying } \int \d z \wedge \bar \delta^{(n)}(z) f(z, \bar z) = (-1)^n n! \left(\frac{\partial^n f}{\partial z^n}\right)_{z=0},
\end{equation}
sampling the nth order zero. The form direction on $\bar \delta^{(n)}([\mu \bar \lambda])$ is in the $\mu$ direction, and $\bar \delta^{(n)}([\mu \bar \lambda])$ is to be interpreted as sampling the nth order zero in the linear combination of $\mu^{\dot 0},\mu^{\dot 1}$ defined by $[\mu \bar \lambda]$ of the integrand.
}.
The $h$ gauge transformation controlled by the gauge parameter $\tilde \eta$ is only an infinitesimal symmetry and by rights we should perform the gauge transformation and field redefinition in an iterative limiting procedure. Taking the gauge transformation all at once will agree with the iterative procedure only at first order in $\tilde \eta$, as we might expect higher order terms from each infinitesimal iteration of the nonlinear gauge transformation. However, the fact that each $\tilde \eta$ comes with a $\bar \delta^{(k-1)}([\mu \bar \lambda])$ in the field redefinition \eqref{eq:field_redef} means that $\tilde \eta$ can only appear linearly in any term in the action due to the fact that $\bar \delta^{(k-1)}([\mu \bar \lambda]) \wedge \bar \delta^{(k'-1)}([\mu \bar \lambda]) \propto \la \lambda \d \bar \mu \ra \wedge \la \lambda \d \bar \mu \ra = 0$.
Therefore we are permitted to take the transformation all at once. The same can be said for the $\tilde h$ gauge transformation controlled by $\tilde \phi$, except that in that case that really is a non-infinitesimal exact symmetry, so we were allowed to do that from the start.

Note that under the initial large gauge transformation of $h$, the variation in the counter term exactly cancels the variation in the path integral measure, and there is no $\tilde \eta$ dependence in the counter term. After the field redefinition \eqref{eq:field_redef}, $h$ in the counter term is shorthand for the combination of the redefined $h'$ and other non-localising terms containing $\tilde \eta$. The counter term will not localise and just come along for the ride in the rest of the paper.

The second term in $S_{\text{SDG}'}$ separates the terms that contain explicit localisation to the locus $[\mu\bar\lambda]^k=0$. 
As we summarized in figure \ref{fig:reduction}, a $\mathbb{C}^*$-scaling reduction to $\mathbb{CP}^1_{\lambda}\times\mathbb{CP}^1_{\mu}\cong\mathbb{MT}\oplus\mathbb{CP}^1_{[\mu\bar\lambda]=0}$ amounts to a further mode expansion in the basis $B_{\Delta}$ \eqref{eq_B_basis} with $\Delta\in\mathbb{Z}+\im\mathbb{R}$ for all the off-shell fields. The action becomes 
\begin{align}
    S_{\text{split}}&= \sum_{k=1}^{\infty}\int_{\mathbb{Z}+\im\mathbb{R}}\d^2\Delta\int_{\mathbb{MT}}\D\lambda\D\mu q^{k+1}\la\lambda\hat\lambda\ra^k\left(\tilde h^M \bar\partial \phi + \eta \bar\partial h^M \right.\nonumber\\
    &\left.+h^M \left(\{ \tilde h^M,\phi\}+\{h^M,\eta\}-\Delta\tilde h^M\right)+\tilde h\left(\{h^M,\phi\}-(\Delta-2\im)h^M\right)\right)+S_{\text{ct-3d}}\nonumber\\
    &+\sum_{k=1}^{\infty}\int_{\mathbb{Z}+\im\mathbb{R}}\d^2\Delta\int\D\lambda\D\mu \wedge \bar\delta^{(k-1)}([\mu \bar\lambda])\left(\tilde \phi \bar\partial \phi + \eta\bar\partial\tilde\eta \right.\nonumber\\
    &\left.+h^{M'} \left(\{ \tilde\phi,\phi\}+\{\tilde\eta,\eta\}-\Delta\tilde\phi\right)+\tilde h^{M'}\left(\{\tilde\eta,\phi\}-(\Delta-2\im)\tilde\eta\right)\right)\,,\label{eq:split_action}
\end{align}
where we have suppressed subscripts to simplify the expressions, we have also slightly abused notation by labelling the minitwistor representative as $h^{M}$ and $\tilde h^{M}$, their localised values are correspondingly $h^{M'}$ and $\tilde h^{M'}$. The first part of the action is a minitwistor action that contains 3d Einstein-Weyl gravity on $\mathbb{H}_3/LdS_3$ slices of Minkowski space\footnote{By turning off $\Delta\neq0$ terms, we recover the scaling reduction of 4d self-dual Einstein-gravity which is known to be 3d Einstein-Weyl gravity \cite{Jones:1985pla,Calderbank98einstein-weylgeometry,Hitchin:1982gh,Hitchin:1983ay}.}. The second part of the action formally localises to the diagonal sphere $\mathbb{CP}^1_{[\mu\bar\lambda]=0}$ with the explicit holomorphic delta function\footnote{For more detail on the meaning of the curly braces, see equation \ref{eq:mod_J}.}. It symbolises compensation response of the entire path integral towards the addition of arbitrarily singular modes in bulk graviton field configurations. After geometrically performing Mellin transform on all the fields, we have obtained this additional piece of the action which lives on the boundary:
\begin{multline}\label{eq:2d_CFT}
    S_{\text{SD}} = \sum_{k=1}^{\infty}\int_{\mathbb{Z}+\im\mathbb{R}}\d^2\Delta\int\D\lambda\D\mu \wedge \bar\delta^{(k-1)}([\mu \bar\lambda])\\
    \boxed{\left(\tilde \phi \bar\partial \phi + \eta\bar\partial\tilde\eta+h^{M'} \left(\{ \tilde\phi,\phi\}+\{\tilde\eta,\eta\}-\Delta\tilde\phi\right)+\tilde h^{M'}\left(\{\tilde\eta,\phi\}-(\Delta-2\im)\tilde\eta\right)\right)}\,.
\end{multline}
Note that the curly braces are not Poisson brackets but rather bilinears of mode numbers $\Delta,k$ for each field in the curly braces, explained in \ref{eq:mod_J}. In the Yang-Mills case \cite{Bu:2023vjt}, we straightforwardly recognized the cubic terms as a coupling between the currents (formed of bilinears in the 2d system scalars) and bulk fields, supported only on the $[\mu \bar \lambda]=0$ locus. Here in gravity, because of the presence of the $\mu^{\dal}$ derivatives in the coupling terms, it is not true that these terms straightforwardly describe a current-bulk field coupling term supported on the locus $[\mu\bar\lambda]=0$.
The $\mu^{\dot \alpha}$ derivatives do not preserve the condition $[\mu \bar \lambda]=0$, and therefore interrogate the extension of the bulk fields off the locus $\{[\mu \bar \lambda]=0\}$.

Bulk field configurations to sum over are Fourier modes or along the polynomial basis we have chosen, which can be naturally interpreted geometrically as extracting pure diffeomorphism modes smeared all over the null boundary. The analogous statement in AdS/CFT would be the coupling of bulk metric perturbation with boundary stress tensor defined over the entire boundary manifold. This is the case here as well, a bulk graviton expanded in modes living on different order formal neighborhoods of $\mathbb{CP}^1_{u=0}$ couples to corresponding mode currents on each order of the neighborhood of the celestial sphere. When one would like to consider the bulk graviton scattering state, the Fourier wavefunction would couple to all of the mode currents. Geometrically, this amounts to viewing bulk graviton as coupling to currents smeared along $\scri$, manifesting the nature of the derivatives we have in the twistor action. 

In the meantime, large pure diffeomorphism generators $\boldsymbol{\tilde{\eta}}(x,k^{\mu})$ and $\boldsymbol{\tilde{\phi}}(x,k^{\mu})$ together with their twistor space counterparts can be expanded in any suitable basis. In particular, even ones which are have high order singular behaviour on the boundary/defect locus. As they label the boundary value the bulk graviton asymptotes to. On twistor space, the uplifted profiles $\tilde\eta$ and $\tilde\phi$ indicates values of the graviton $H^{0,1}$ representatives pulled back to the particular sphere $\mathbb{CP}^1_{[\mu\bar\lambda]=0}$. For instance, one could start with boundary conditions where all fields together with their gauge transformations vanish on $\mathbb{CP}^1_{[\mu\bar\lambda]=0}$, via a singular gauge transformation that does not vanish on the sphere, one could obtain graviton configurations which pulls back to singular profiles. Note that since such gauge transformations are trivial almost everywhere else, we still have the same split theory at hand as long as we also sum over all such defect field configurations\footnote{This is a classical statement, on the level of path integrals, this is only true provided that the original Lagrangian theory does not suffer from gauge anomaly.}.

\paragraph{Mode currents:}
Zooming in on the coupling terms in \eqref{eq:2d_CFT}, one can integrate by parts on the derivatives of the holomorphic delta function which results in a sum over all possible distributions of the derivatives along $[\mu\bar\lambda]$ on the bulk and individual component fields. Due to the presence of the sum over all positive integer values of $k$, we can choose to expand all the fields on the following complete basis along $[\mu\bar\lambda]$ and the other component of $\mu^{\dal}$: $[\mu\bar\lambda]/\la\lambda\hat{\lambda}\ra$, $[\mu\hat{\bar\lambda}]$. Then a basis function can be written as 
\begin{equation}
    Y_{\Delta,k} = q^\Delta\,u^k\,,
\end{equation}
where $\Delta\in\mathbb{Z}+\im\mathbb{R}$ and $k\in\mathbb{Z}_+$\footnote{It looks like that we are only considering basis where $u$ and $q$ are holomorphic in $\mu^{\dal}$, in fact, we are suppressing the antiholomorphic $\bar u$ and $\bar q$ dependences as they are ignorant about the Poisson $\mu^{\dal}$ derivatives. The antiholomorphic weights gets fixed to $1$ after integrating against the holomorphic delta function.}. One perspective one could take is to look at the currents that couple to individual modes of the gravitons $h_{\Delta,k}$, $\tilde h_{\Delta,k}$, which will have two labels. $\bar \delta^{(k-1)}([\mu\bar \lambda])$ samples the $(k-1)$th order zero in $[\mu \bar \lambda]$ meaning that untilded fields in the kinetic term appear strictly with negative $k$ (since in our convention, $k$ is the order of the singularity at $[\mu \bar \lambda]=0$). Untilded fields with postive $k$ are auxillary on the $\mathbb{CP}^1_{[\mu \bar \lambda]=0}$ and do not propagate, but do appear in the cubic interaction terms.
\begin{multline}
    S_{\text{SD}} = \sum_{k\in\mathbb{Z}_+}\int_{\mathbb{Z}+\im\mathbb{R}}\d^2\Delta\int_{\mathbb{CP}^1_{[\mu\bar\lambda]=0}}\D\lambda\, \left(\tilde \phi \bar\partial \phi + \eta\bar\partial\tilde\eta \right.\nonumber\\\left.+h^{M'} \underbrace{\left(\{ \tilde\phi,\phi\}+\{\tilde\eta,\eta\}-\Delta\tilde\phi\right)}_{J_{\Delta,k}}+\tilde h^{M'}\underbrace{\left(\{\tilde\eta,\phi\}-(\Delta-2\im)\tilde\eta\right)}_{\tilde J_{\Delta,k}}\right)\,,
\end{multline}
where the kinetic terms now give the following OPE
\begin{align}
    &\tilde\phi_{\Delta_1,k_1}(\lambda_1)\phi_{\Delta_2,k_2}(\lambda_2)\sim \frac{\delta_{k_1+k_2,0}\delta_{\Delta_1+\Delta_2,0}}{\la 12\ra}\left(\frac{\la 1 \iota \ra}{\la 2 \iota \ra}\right)^{1- 2 \text{Re}(\Delta_1)}\,, \nonumber
    \\
    &\tilde\eta_{\Delta_1,k_1}(\lambda_1)\eta_{\Delta_2,k_2}(\lambda_2)\sim \frac{\delta_{k_1+k_2,0}\delta_{\Delta_1+\Delta_2,0}}{\la 12\ra}\left(\frac{\la 1 \iota \ra}{\la 2 \iota \ra}\right)^{1- 2 \text{Re}(\Delta_1)}\,. \label{contraction_mode}
\end{align}
$\iota^{\alpha}$ is the reference spinor associated to a choice of gauge in the propagator that arises from inverting the $\bar \partial$ operator. The extra factors $2\text{Re}(\Delta_1)$ of the reference spinors are there because the fields with conformal weight $\Delta_1\in\mathbb{Z}+\im\mathbb{R}$ have weight $\mathcal{O}(-2\text{Re}(\Delta_1))$. 
In terms of local coordinates $\lambda^{\alpha}=(1,\, z)$, $\bar\lambda^{\dal}= (1,\,\bar z)$ on the $\mathbb{C}$ patch of $\mathbb{CP}^1$ where we exclude the point $\lambda^{\alpha} \propto \iota^{\alpha}$, the OPEs take the familiar form:
\begin{align}\label{contraction_mode_local}
    &\tilde\phi_{k_1,\Delta_1}(z_1)\phi_{k_2,\Delta_2}(z_2)\sim \frac{\delta_{k_1+k_2,0}\delta_{\Delta_1+\Delta_2,0}}{z_1-z_2}\,, \quad \tilde\eta_{k_1,\Delta_1}(z_1)\eta_{k_2,\Delta_2}(z_2)\sim \frac{\delta_{k_1+k_2,0}\delta_{\Delta_1+\Delta_2,0}}{z_1-z_2}\,.
\end{align}
And the mode currents $J_{\Delta,k}$ and $\tilde J_{\Delta,k}$ coupling to modes of the bulk graviton can be read off from the action:
\begin{multline}
    J_{\Delta,k} = \sum_{k_1,k_2\in\mathbb{Z}_+}\int\d^2\Delta_1\d^2\Delta_2\,\delta_{k_1+k_2,k}\delta_{\Delta_1+\Delta_2,\Delta}\,(\tilde\phi_{\Delta_1,k_1}\phi_{\Delta_2,k_2}+\tilde\eta_{\Delta_1,k_1}\eta_{\Delta_2,k_2}) \\
    \left(\partial_{\mu^{\dal}}Y_{\Delta_1,k_1} \partial_{\mu_{\dal}}Y_{\Delta_2,k_2}+\partial_{\mu^{\dal}}Y_{\Delta_1,k_1} \partial_{\mu_{\dal}}Y_{\Delta_2,k_2}\right)-\Delta\tilde\phi_{k,\Delta}\,,
\end{multline}
where we need to compute the result of applying the Poisson bivector to a pair of basis functions. Note that in order for the mode numbers to make sense as part of the edge action $S_{\text{edge}}$, it was important to project $\mu^{\dal}$ onto the basis $\bar\lambda^{\dal}/\la\lambda\hat{\lambda}\ra$ and $\hat{\bar\lambda}^{\dal}$. Suppressing the label numbers, this allows us to write down the mode currents:
\begin{equation}\label{eq:mod_J}
    J_{\Delta,k} = \sumint_{k_1,k_2,\Delta_1,\Delta_2} (\Delta_2k_1-\Delta_1k_2)(\tilde\phi\phi+\tilde\eta\eta)\,\delta_{k_1+k_2,k}\delta_{\Delta_1+\Delta_2,\Delta}-\Delta\tilde\phi_{k,\Delta}\,.
\end{equation}
We shall see that given the contraction rule between free fields in the action, the OPE between the mode currents can be computed, it is perhaps not surprising that we shall find the celebrated $\cL w_{1+\infty}$ algebra. Similarly, the mode current coupling to negative helicity graviton can be read off 
\begin{equation}\label{eq:mod_tJ}
    \tilde J_{\Delta, k}=\sumint_{k_1,k_2,\Delta_1,\Delta_2} (\Delta_2k_1-\Delta_1k_2)\tilde\eta\phi\,\delta_{k_1+k_2,k}\delta_{\Delta_1+\Delta_2,\Delta}-(\Delta-2\im)\tilde\eta_{k,\Delta}\,.
\end{equation}

\paragraph{Fourier currents:}
A second perspective one can take is to look at the mode currents collectively. This amounts to formally not evaluating the delta function to avoid focusing on individual $u$ modes of the free fields, by expanding the bulk graviton in Fourier modes while keeping the $\Delta$ labels on the field from the $\mathbb{C}^*$-scaling reduction, we can recognize the Fourier currents coupling to graviton scattering states. We find it cleanest to begin by rewinding the geometric Mellin transform (the scaling reduction) in the 2d action \eqref{eq:2d_CFT}:
\begin{multline}
    \sum_{k\in\mathbb{Z}_+}\int_{\mathbb{PT}^\cO}\D^3Z\wedge \bar\delta^{(k-1)}([\mu\bar\lambda])\\\left(\int\d^4p \e^{\im p\cdot x}\,h'(\lambda,p^{\alpha\dal})\left(\{\tilde\phi,h'\}+\{\tilde\eta,\tilde h'\}\right)+\int\d^4p \e^{\im p\cdot x}\,\tilde h'(\lambda,p^{\alpha\dal})\{\tilde\eta,h'\}\right)
\end{multline}
where we have expanded the bulk graviton modes in Fourier modes. Then the $\mathbb{C}^{*}$-scaling reduction on the rest of the fields gives 
\begin{multline}
    \int\D\lambda\wedge\D\mu\wedge \bar\delta^{(k-1)}([\mu\bar\lambda])\\
    \int\d^4p \,h^{M'}(\lambda,p^{\alpha\dal})\underbrace{\int \frac{\d q \d \bar q}{q \bar q}\sum_{k\in\mathbb{Z}_+}\e^{\im p\cdot x}\int\d^2\Delta B_\Delta \left(\{\tilde\phi,\phi\}+\{\tilde\eta,\eta\}-\Delta\tilde\phi\right)}_{J(\lambda,p^{\alpha\dal})}\\
    +\int\d^4p \,\tilde h^{M'}(\lambda,p^{\alpha\dal})\underbrace{\int \frac{\d q \d \bar q}{q \bar q}\sum_{k\in\mathbb{Z}_+}\e^{\im p\cdot x}\int\d^2\Delta B_\Delta\left(\{\tilde\eta,\phi\}-(\Delta-2\im)\tilde\eta\right)}_{J(\lambda,p^{\alpha\dal})}
\end{multline}
The basis function $B_{\Delta}$ \eqref{eq_B_basis} is what remains after having done the mode decomposition in the $q = [\mu \hat{\bar \lambda}]$, trading $q, \bar q$ for the mode labels $\Delta$.
As usual, we have hidden the $\Delta$ labels on all the fields. This allows us to define the collective current $J\in\Omega^{1,0}\otimes\cO(2)$ which essentially amounts to resumming all the mode currents in the $k$ basis along $\scri$. The collective current can be regarded as a smeared object on all higher order neighborhood of the light cone cut at $u=0$, which couples to bulk graviton scattering states.
\begin{equation}\label{Fourier_currentJ}
J(\lambda,p^{\alpha\dal})=\D\lambda\int \frac{\d q \d \bar q}{q \bar q}\sum_{k\in\mathbb{Z}_+}\int \d^2\Delta \int\D\mu\wedge\bar\delta^{k-1}([\mu\bar\lambda])B_{\Delta}\e^{\im p\cdot x}\,\left(\{\phi,\tilde\phi\}+\{\eta,\tilde\eta\}-\Delta\tilde\phi
    \right)\,.
\end{equation}
where we shall use $[\mu \bar\lambda]\frac{\la \hat \lambda p \hat{\bar\lambda}]}{\la\lambda \hat\lambda\ra^2}+\text{c.c.}=\im p\cdot x$ to expand Fourier basis in the 2d coordinates in the OPE computations later. Similarly, the Fourier current coupling to negative helicity scattering states can be read off from the action 
\begin{equation}\label{Fourier_current_tJ}
    \tilde J(\lambda,p^{\alpha\dal})= \D\lambda\int \frac{\d q \d \bar q}{q \bar q} \sum_{k\in\mathbb{Z}_+}\int\d^2\Delta\,\D\mu\wedge\bar\delta^{k-1}([\mu\bar\lambda])\,B_{\Delta} \e^{\im p\cdot x}\,\left(\{\tilde\eta,\phi\}-(\Delta-2\im)\tilde\eta\right)\,.
\end{equation}
These Fourier currents describe the collective coupling behaviour of the boundary 2d CFT to graviton scattering states, unlike the mode currents from expanding free fields on the orthogonal basis, which reflect interactions between graviton modes localised to individual order neighborhood of the celestial sphere, the dynamics of Fourier currents intuitively computes interaction of bulk graviton scattering states. We shall demonstrate a correlation function computations in the presence of background interaction vertices. 

\section{The 2d CFT}\label{sec:OPEs}
In this section, we further study the 2d chiral CFT \eqref{eq:2d_CFT} we have obtained after properly including boost eigenstates in the field configuration and geometrically Mellin transform the action functionals. 
\subsection{Mode current OPE}\label{subsec:mode_OPE}
Using the definitions of the mode currents \eqref{eq:mod_J}, \eqref{eq:mod_tJ}, we can compute their OPEs using holomorphic free field contractions rules \eqref{contraction_mode}. For example, we can take contractions between two $J_{\Delta,k}$s, as a demonstration, we have only written out the calculation for the $\phi\tilde\phi$ part, with the $\eta\tilde\eta$ part following in a straightforward fashion. First we take single contractions which results in the following two terms 
\begin{multline}
    J_{\Delta,k}(z_1)J_{\Delta',k'}(z_2)\sim\sumint_{k_i,k_i',\Delta_i,\Delta_i'} \delta_{\Delta,\Delta_1+\Delta_2}\delta_{k,k_1+k_2}\delta_{\Delta',\Delta'_1+\Delta'_2}\delta_{k',k'_1+k'_2}(k_1\Delta_2-k_2\Delta_1)\\
    \left(\delta_{k_1+k'_2,0}\delta_{\Delta_1+\Delta'_2,0}\tilde\phi_{k_2,\Delta_2}\phi_{k'_1,\Delta'_1}-\delta_{k_2+k'_1,0}\delta_{\Delta_2+\Delta'_1,0}\phi_{k_1,\Delta_1}\tilde\phi_{k'_2,\Delta'_2}\right)\,\frac{(k'_1\Delta'_2-k'_2\Delta'_1)}{z_{12}}\,,
\end{multline}
using individual delta functions, one could eliminate the summation or integral over $k'_i$ and $\Delta'_i$ in each term
\begin{multline}
   \sim\frac{1}{z_{12}}\sumint_{k_i,\Delta_i} \delta_{\Delta,\Delta_1+\Delta_2}\delta_{k,k_1+k_2}\left((k_1\Delta_2-k_2\Delta_1)\left(k_1(\Delta'+\Delta_1)-\Delta_1(k'+k_1)\right)\tilde\phi_{k_2,\Delta_2}\phi_{k'+k_1,\Delta'+\Delta_1}\right.\\\left.+(k_1\Delta_2-k_2\Delta_1)\left(\Delta_2(k'+k_2)-k_2(\Delta'+\Delta_2)\right)\tilde\phi_{k'+k_2,\Delta'+\Delta_2}\phi_{k_1,\Delta_1}\right)\,,
\end{multline}
after shifting $k_1\to k_1+k'$, $k_2\to k_2-k'$, $\Delta_1\to\Delta'+\Delta_1$ and $\Delta_2\to\Delta_2-\Delta'$ in the second term, we can combine the two terms and have a unified expression for the simple pole
\begin{multline}
    J_{\Delta,k}(z_1)J_{\Delta',k'}(z_2)\sim 
    \sumint_{k_i,\Delta_i}\delta_{\Delta,\Delta_1+\Delta_2}\delta_{k,k_1+k_2}\\
    \left(\Delta k'-\Delta' k\right)\left((k'+k_1)\Delta_2-(\Delta_1+\Delta')k_2 \right)\,\frac{\tilde\phi_{k_2,\Delta_2}\phi_{k_1+k',\Delta'+\Delta_1}}{z_{12}}\,,
\end{multline}
where we can recognize the last part recombines into a mode current $J_{\Delta+\Delta',k+k'}$ with mode summed up. Overall, keeping all the possible Wick contractions in mind, we could summarize the expression for the simple pole in the OPE between mode currents:
\begin{equation}
    \boxed{J_{\Delta,k}(z_1)J_{\Delta',k'}(z_2)\sim \frac{\Delta k'-\Delta' k}{z_{12}}\,J_{\Delta+\Delta',k+k'}(z_2)} \,,
\end{equation}
where we have turned off the double poles by restricting ourselves to the subsector $\Delta+\Delta'\neq 0$, $k+k'\neq 0$. The simple pole term is just the famous $\cL w_{1+\infty}$ algebra if we were to shift $k\to k-1$ and $k'\to k'-1$:
\begin{equation}
    J_{\Delta,k-1}(z_1)J_{\Delta',k'-1}(z_2)\sim\frac{\Delta (k'-1)-\Delta'(k-1)}{z_{12}}\,J_{\Delta+\Delta',k+k'-2}(z_2)\,.
\end{equation}
The shift in $k$ to align with the conventional mode numbers suggests that it is perhaps more natural to label the poles in the large gauge transformations $\frac{1}{[\mu \bar \lambda]^{1+k}}$, such that the case corresponding to $k=0$ is the simple pole. A crucial difference with the usual $\cL w_{1+\infty}$ is the absence of the wedge condition $-k\leq \Delta\leq k$, in contrast there are no restrictions for the algebra we have obtained\footnote{Hence it is the loop algebra of the $w_{1+\infty}$ rather than the loop algebra of the wedge subalgebra of $w_{1+\infty}$.}. The way our currents are organized is reminiscent of the construction on $\scri$ in terms of News tensor and its $u$-derivatives in \cite{Donnay:2024qwq}.

Unlike the usual $\cL w_{1+\infty}$ algebra, the first mode number is not $\m \in \mathbb{Z}$ but rather $\Delta = m + \im s \in \mathbb{Z} + \im \mathbb{R}$. The additional continuous mode number $s$ is associated to the $\mathbb{R}^+$ scaling reduction that we have performed on spacetime, which is not usually performed in the derivation of the $\cL w_{1+\infty}$ algebra from twistor space.

One is also allowed to take double contractions, giving the following double pole
\begin{multline}
    J_{\Delta,k}(z_1)J_{\Delta',k'}(z_2)\sim\sumint_{k_i,k_i',\Delta_i,\Delta_i'} \delta_{\Delta,\Delta_1+\Delta_2}\delta_{k,k_1+k_2}\delta_{\Delta',\Delta'_1+\Delta'_2}\delta_{k',k'_1+k'_2}\\
    \frac{(k_1\Delta_2-k_2\Delta_1)(k'_1\Delta'_2-k'_2\Delta'_1)\delta_{k_1+k'_2,0}\delta_{k_2+k'_1,0}\delta_{\Delta_1+\Delta'_2,0}\delta_{\Delta_2+\Delta'_1,0}}{z_{12}^2}\,.
\end{multline}
After evaluating the delta functions on the mode numbers, the double pole contribution becomes
\begin{equation}\label{mode_JJ_double_pole}
    J_{\Delta,k}(z_1)J_{\Delta',k'}(z_2)\sim -\frac{\delta_{\Delta+\Delta',0}\delta_{k+k',0}}{z_{12}^2}\,\sumint_{k_i,\Delta_i} \delta_{\Delta,\Delta_1+\Delta_2}\delta_{k,k_1+k_2}
    (k_1\Delta_2-k_2\Delta_1)^2\,.
\end{equation}
Note the presence of the sum over $\Delta_i\in\mathbb{Z}+\im\mathbb{R}$ and $k_i\in\mathbb{Z}_+$, which diverges. In order for such "level" term to vanish, we shall always restrict ourselves to the subsector of the modes where
\begin{equation}
    \Delta+\Delta'\neq 0\,,\quad k+k'\neq 0\,,
\end{equation}
which also happens to be the subsector where the simple pole term is non-trivial (not always ending up with the zero mode $J_{0,0}$). This allows us to explicitly recover the algebra the mode currents built out of large pure diffeomorphism generator smeared over $\scri$ and bulk graviton scalar components. Similarly, one can also compute the contractions between $J_{k,\Delta}$ and $\tilde J_{k,\Delta}$ or a pair of $\tilde J_{k,\Delta}$s. In particular, it is not difficult to see that given the contraction rules we have in the 2d CFT, the latter is automatically $0$. We summarize the results here
\begin{equation}
    \begin{split}
         &J_{\Delta,k}(z_1)J_{\Delta',k'}(z_2)\sim\frac{\Delta k'-\Delta' k}{z_{12}}\,J_{\Delta+\Delta',k+k'}(z_2)\,;\\
         &J_{\Delta,k}(z_1)\tilde J_{\Delta',k'}(z_2)\sim\frac{\Delta k'-\Delta' k}{z_{12}}\,\tilde J_{\Delta+\Delta',k+k'}(z_2)\,;\\
         &\tilde J_{\Delta,k}(z_1)\tilde J_{\Delta',k'}(z_2)\sim 0\,.
    \end{split}
\end{equation}
$J_{0,0}$ is the unique central element in the algebra generated by $J_{\Delta,k}$, but unfortunately it vanishes. We can take the following group contraction, however, and write the following rescaled $J_{0,0}$ which is nonzero and remains central (this will correspond to the generator of the Cartan subalgebra of our extension of the $\cL w_{1+\infty}$ algebra to noninteger mode numbers), using the expression for $J_{\Delta,k}$ in \eqref{eq:mod_J}:
\begin{equation}
    \lim_{\epsilon \rightarrow 0} \frac{J_{\epsilon,0}}{\epsilon} = \int \d^2\Delta \sum_p p\left(\phi_{-\Delta,-p} \tilde \phi_{\Delta, p} + \eta_{\Delta, p} \tilde\eta_{-\Delta,-p} \right) + \lim_{\epsilon\rightarrow 0}\tilde \phi_{\epsilon,0}\,,
\end{equation}
the linear term vanishes because there is no $\tilde \phi$ at $k=0$ (our gauge transformations only introduce tilded fields with $k\in\mathbb{Z}_+$). This current measures the quantised U(1) charge $k \in \mathbb{Z}$, being the strength of the zero/singularity of the product of bulk fields at $[\mu \bar \lambda]$:
\begin{equation}
    \lim_{\epsilon \rightarrow 0} \frac{J_{\epsilon,0}}{\epsilon} J_{\Delta, k} \sim\lim_{\epsilon\to 0}\frac{\epsilon k}{\epsilon\,z_{12}}\,J_{\Delta+\epsilon,k}= \frac{k}{z_{12}} J_{\Delta, k}\,.
\end{equation}
We can build a U(1) charge from the current (with a smooth function $f$):
\begin{equation}
    Q(f):=-\int_{\mathbb{CP}^1} \d z \, \bar \partial f \,\, \lim_{\epsilon \rightarrow 0} \frac{J_{\epsilon,0}}{\epsilon}\,.
\end{equation}
This is a cohomological gauge transformation in the U(1) connection $h_{\epsilon,0} \rightarrow h_{\epsilon,0} + \bar \partial f_{\epsilon,0}/\epsilon$. It is unsurprising and can be quickly checked that the exponentiated charge acts as a U(1) transformation on the mode currents:
\begin{equation}
    e^{\im Q(f)} J_{\Delta, k}(z) = e^{\im k f(z)}\, J_{\Delta, k}(z)\,.
\end{equation}
However, note that this is not a symmetry of the chiral 2d CFT action \eqref{eq:2d_CFT} in general. Consider the coupling term $\int \D \lambda \,h^{M'}J $. Under such a transformation, we instead have:
\begin{equation}
      \e^{\im Q(f)} \left(h^{M'}_{-\Delta, -k} J_{\Delta, k}\right)(z) \rightarrow \left(h^{M'}_{-\Delta, -k} \e^{\im k f(z)} J_{\Delta, k}\right)(z)\,.
\end{equation}
In order for this to be a symmetry, $h^{M'}_{-\Delta, -k}$, the boundary value of the bulk graviton mode, must be rotated as well by $\e^{-\im k f(z)}$. This is a (gauge) motion in the space of boundary conditions for the theory, and closely related to the supertranslations studied in \cite{Himwich:2020rro}.

Consider a setting in which $h$ is an asymptotically free on-shell graviton state satisfying the free equation of motion on $\mathbb{PT}$: $\bar\partial h = 0$. We recognize $h$ as the negative helicity radiative degree of freedom. Since $h \in \Omega^{0,1}(\mathbb{CP}^1,\mathcal{O}(2))$, $\bar\partial$-closure means it is $\bar\partial$-exact. Define $C := \bar\partial^{-1} \lim_{\epsilon\rightarrow 0} \epsilon  h_{-\epsilon,0}$.
If we were to consider the following "dressed" bulk states:
\begin{equation}\label{dressed_op}
    \e^{ikC}\mathcal{O}_{\Delta,k}
\end{equation}
of some bulk particle state $\mathcal{O}_{\Delta,k}$. This is a graviton in our case and an electron in QED. As defined above, a U(1) gauge transformation takes
\begin{equation}
     C \rightarrow C - f\,.
\end{equation}
And therefore the dressed state is now U(1) gauge invariant. Such dressed bulk states \eqref{dressed_op} is the Wilson line dressing first discussed in the context of QED \cite{Kulish:1970ut,Nguyen:2023ibj} and then in gravity \cite{Himwich:2020rro}, where $C$ was identified as the Goldstone associated with the spontaneous breaking of supertranslations. For us, it is related to $(\bar\partial^{-1}h)_{0,0}$, where $(\Delta,k)={(0,0)}$ means that it is the part that is non-singular on the celestial sphere and homogenous of degree 0 in $x$, i.e. the large pure diffeomorphism profile regular everywhere on $\scri$.


This manifests the way our boundary currents changes the boundary conditions for bulk gravitons, which provide explicit demonstration of how the correspondence between bulk gravity and this chiral 2d CFT \eqref{eq:2d_CFT} works. Instead of studying 4d S-matrices, in our formalism considerations of charges and large gauge transformations such as these are immediate and totally explicit, with OPEs and couplings following from a totally explicit 2d CFT Lagrangian formulation. We will revisit further applications of our model in a future paper.

\subsection{Full current OPE}\label{subsec:full_OPE}
As promised, we also derive the OPEs between Fourier currents \eqref{Fourier_currentJ} and \eqref{Fourier_current_tJ} with OPE rules for the free fields smeared over $\scri$ \eqref{contraction_mode}. We hide the integration over $ q$ as they are bystanders throughout the computation. Focusing on the $\phi-\tilde\phi$ part of the currents\footnote{It works exactly the same for the $\tilde\eta-\eta$ system.}:
\begin{equation}
\begin{split}
&J(\lambda_1,p_1^{\alpha\dal})J(\lambda_2,p_2^{\alpha\dal})\sim \, \int\d^2\Delta\d^2\Delta'\sum_{k_1,k_2}\,\frac{\D\lambda_1\D\lambda_2}{\la 12\ra}\frac{\la\lambda_1\iota\ra\la\lambda_1\upsilon\ra}{\la\lambda_2\iota\ra\la\lambda_2\upsilon\ra}\int\D\mu_1\D\mu_2\wedge\\
   & \bar\delta^{k_1-1}([\mu_1\bar\lambda_1])\wedge\bar\delta^{k_2-1}([\mu_2\bar\lambda_2])\left(\partial_{\mu_1^{\dal}}\left(\e^{\im p_1\cdot x}B_\Delta\right)\partial_{\mu_{2\dbl}}\left(\e^{\im p_2\cdot x}B_{\Delta'}\right) \partial_{\mu_{1\dal}}\phi\partial_{\mu_2^{\dbl}}\tilde\phi+\right.
   \\
   &\left.\partial_{\mu_{1\dal}}\left(\e^{\im p_1\cdot x}B_\Delta\right)\partial_{\mu_2^{\dbl}}\left(\e^{\im p_2\cdot x}B_{\Delta'}\right) \partial_{\mu_1^{\dal}}\tilde\phi\partial_{\mu_{2\dbl}}\phi\right)\,,\label{nasty_2}
\end{split}
\end{equation}
where we have integrated by parts for the $\mu^{\dal}$ derivatives acting on the free fields we are taking the contractions, note that the derivatives hitting the holomorphic delta functions producing current modes localised on higher order formal neighborhoods of the sphere. This has been accounted for by the infinite sum of all higher orders neighborhoods in the definition of the currents\footnote{The extra numerical factors the derivatives bring down are functions of $k_1$ $k_2$, exactly as in \eqref{eq:nasty_detail}, which contribute numerical factors that can be reabsorbed into redefinitions of the currents. Complete presentation of the computation would be unenlightening, in order to have a comprehensible discussion, we suppress those derivatives in \eqref{nasty_2}.}. Physically, this suggests that the current coupling to a bulk graviton Fourier mode is not localised on $\scri$, but rather smeared over entirety of the null boundary (see Appendix \ref{Appendix:C}). The two terms Schouten to combine into one term. In addition, we note the $\mu^{\dal}$ derivatives acting on Fourier basis gives
\begin{equation}
    \partial_{\mu^{\dal}}\e^{\im p\cdot x} = \frac{p_{\alpha\dal}\hat{\lambda}^{\alpha}}{\la\lambda\hat{\lambda}\ra}\,\e^{\im p\cdot x}\,.
\end{equation}
and its action on the basis $B_{\Delta}$ would pull down a factor of $\Delta$ with the holomorphic weight in $B_{\Delta}$ reduced by $1$, we shall write it as 
\begin{equation}
    B_{h,\bar h} = u^h \bar u^{\bar h} ,\quad \partial_{\mu^{\dal}} B_{h,\bar h}= \frac{\bar\lambda_{\dal}}{\la\lambda\hat{\lambda}\ra} B_{h-1,\bar h}\,.
\end{equation}
where $h$ and $\bar h$ are the holomorphic and antiholomorphic weights in $u$ and $\bar u$\footnote{$h=\Delta$ and $\bar h = -\bar \Delta$ from the definition of $B_{\Delta}$ in \eqref{eq_B_basis}.}. Evaluating the derivatives gives
\begin{multline}\label{eq:nasty_detail}
    \int\d h\d\bar h \d h'\d\bar h' \frac{\D\lambda_2\,[\bar\lambda_1\bar\lambda_2]}{\la 12\ra\la\lambda_1\hat{\lambda}_1\ra\la\lambda_2\hat{\lambda}_2\ra}\frac{\la\lambda_1\iota\ra\la\lambda_1\upsilon\ra}{\la\lambda_2\iota\ra\la\lambda_2\upsilon\ra}\,\e^{\im(p_1+p_2)\cdot x}\\
    \left(B_{h,\bar h}B_{h',\bar h'}+h'B_{h,\bar h}B_{h'-1,\bar h'}+hB_{h-1,\bar h}B_{h',\bar h'}+hh'B_{h-1,\bar h}B_{h'-1,\bar h'}\right)\delta_{h+h',0}\delta_{\bar h+\bar h',0}\{\phi,\tilde\phi\}\,,
\end{multline}
where we have made the choice $p^{\alpha\dal}=\frac{\lambda^{\alpha}\bar\lambda^{\dal}}{\la\lambda\hat{\lambda}\ra}$. The integrals on the basis can be done explicitly using the orthogonality relations of the basis giving us some number which can be absorbed into redefinitions of the definitions of the Fourier currents. Summarizing the result, we have:
\begin{equation}
J(\lambda_1,p_1^{\alpha\dal})J(\lambda_2,p_2^{\alpha\dal})\sim \frac{\D\lambda_2\,[\bar\lambda_1\bar\lambda_2]}{\la 12\ra\la\lambda_1\hat{\lambda}_1\ra\la\lambda_2\hat{\lambda}_2\ra}\frac{\la\lambda_1\iota\ra\la\lambda_1\upsilon\ra}{\la\lambda_2\iota\ra\la\lambda_2\upsilon\ra}\,J(\lambda_1,p_1^{\alpha\dal}+p_2^{\alpha\dal})\,,
\end{equation}
Equivalently, we could have evaluated the final current at $\lambda_2$, which contributes another term
\begin{equation}\label{eq:JJ_OPE_full}
J(\lambda_1,p_1^{\alpha\dal})J(\lambda_2,p_2^{\alpha\dal})\sim \frac{1}{2}\frac{\D\lambda_2\,[\bar\lambda_1\bar\lambda_2]}{\la 12\ra\la\lambda_1\hat{\lambda}_1\ra\la\lambda_2\hat{\lambda}_2\ra}\frac{\la\lambda_1\iota\ra\la\lambda_1\upsilon\ra}{\la\lambda_2\iota\ra\la\lambda_2\upsilon\ra}\,J(\lambda_1,p_1^{\alpha\dal}+p_2^{\alpha\dal}) + 1\leftrightarrow 2\,.
\end{equation}
We recover the familiar leading inverse soft factor as discussed in \cite{Bern:1998sv}, diagrammatically, this suggests that taking an OPE on the celestial sphere with our 2d CFT amounts to taking all possible ways to re-attach a soft leg, inside a correlation function, one is required to take all such contractions to obtain higher point functions. We are free to make choices for the reference spinors $\iota^{\alpha}$ and $\upsilon^{\alpha}$, they vanish on the support of momentum conservation of all currents in a correlation function producing different forms of form factors. 

One is also allowed to take double contractions in the OPE, which contributes 
\begin{multline}            
J(\lambda_1,p_1^{\alpha\dal})J(\lambda_2,p_2^{\alpha\dal})\sim\frac{\D\lambda_1\D\lambda_2\,[\bar\lambda_1\bar\lambda_2]^2}{\la12\ra^2\la\lambda_1\hat{\lambda}_1\ra^2\la\lambda_2\hat{\lambda}_2\ra^2}\sum_{k_1,k_2}\int\D\mu_1\D\mu_2\wedge\bar\delta^{k_1-1}([\mu_1\bar\lambda_1])\wedge\bar\delta^{k_2-1}([\mu_2\bar\lambda_2])\times\\
\,\e^{\im(p_1+p_2)\cdot x}\int\d h\d\bar h\d h'\d \bar h'\delta_{h+h',0}\delta_{\bar h+\bar h',0}\,\\\left(hh'B_{h-1,\bar h}B_{h'-1,\bar h'}+h^2h'B_{h-2,\bar h}B_{h'-1,\bar h'}+hh'^2B_{h-1,\bar h}B_{h'-2,\bar h'}+h^2h'^2B_{h-2,\bar h}B_{h'-2,\bar h'}\right)\,.
\end{multline}
We could once again omit the numerical factors from doing the mode number integrals on the basis functions since they could be absorbed into the currents. The interpretation of the double pole here is a bit subtle, integrating by parts on the derivatives on the delta functions, we can get double pole to be
\begin{equation}            
J(\lambda_1,p_1^{\alpha\dal})J(\lambda_2,p_2^{\alpha\dal})\sim\frac{\D\lambda_1\D\lambda_2\,[\bar\lambda_1\bar\lambda_2]^2}{\la12\ra^2\la\lambda_1\hat{\lambda}_1\ra^2\la\lambda_2\hat{\lambda}_2\ra^2}
\sum_{k_1,k_2}\frac{\im^{k_1+k_2+1}}{\la\lambda_1\hat{\lambda}_1\ra^{k_1-1}\la\lambda_2\hat{\lambda}_2\ra^{k_2-1}}\,,
\end{equation}
where the exponential has been set to $1$ on the locus of $[\mu_i\bar\lambda_i]=0=p_i\cdot x$. This double pole is reminiscent of a subscattering process when a Feynman diagram becomes disconnected, signalling self-energy type subleading $\cO(\hbar)$ contributions in the factorisation of an $n$ point process to a 2 and an $(n-2)$-point process, which we shall leave for future studies. For the purpose of this paper, which concerns tree-level computations, we shall focus on the leading simple pole contribution.

Similar computation can be done for other combinations of currents, we summarize the results here
\begin{equation}
    \begin{split}
&J(\lambda_1,p_1^{\alpha\dal})J(\lambda_2,p_2^{\alpha\dal})\sim \frac{\D\lambda_2\,[\bar\lambda_1\bar\lambda_2]}{\la 12\ra\la\lambda_1\hat{\lambda}_1\ra\la\lambda_2\hat{\lambda}_2\ra}\frac{\la\lambda_1\iota\ra\la\lambda_1\upsilon\ra}{\la\lambda_2\iota\ra\la\lambda_2\upsilon\ra}\,J(\lambda_1,p_1^{\alpha\dal}+p_2^{\alpha\dal})\,;\\
        &J(\lambda_1,p_1^{\alpha\dal})\tilde J(\lambda_2,p_2^{\alpha\dal})\sim \frac{\D\lambda_2\,[\bar\lambda_1\bar\lambda_2]}{\la 12\ra\la\lambda_1\hat{\lambda}_1\ra\la\lambda_2\hat{\lambda}_2\ra}\frac{\la\lambda_1\iota\ra\la\lambda_1\upsilon\ra}{\la\lambda_2\iota\ra\la\lambda_2\upsilon\ra}\,\tilde J(\lambda_1,p_1^{\alpha\dal}+p_2^{\alpha\dal})\,;\\
        & \tilde J(\lambda_1,p_1^{\alpha\dal})\tilde J(\lambda_2,p_2^{\alpha\dal})\sim 0\,.
    \end{split}
\end{equation}
Although the OPEs depend on reference spinors, note that the form factors that they generate in the next section will be independent of the choice of reference spinors. This is because form factors are defined on the support of an overall 4d momentum conserving $\delta$ function, and changing the reference spinors will produce terms that vanish on the support of the overall momentum conserving $\delta$ function. In order to simplify correlator computations, we can make particular choices to eliminate entire sets of OPEs to obtain the most convenient form of graviton scattering form factors.

\section{Marginal deformation from SD gravity to full gravity}\label{sec:deformation}
Further from describing the OPEs of the uplifted large pure diffeomorphism generators and their coupling to bulk fields, it would be interesting to compute 2d correlation functions which produces corresponding observables of graviton scattering states in 4d. For example, the simplest example is the 3-point MHV graviton scattering amplitude, which amounts to inserting a 3-point interaction vertex with three free on-shell gravitons propagating to $\scri$. On $\mathbb{PT}^\cO$, this is manifested as insertions of on-shell $H^{0,1}$ wavefunctions on copies of the twistor $\mathbb{CP}^1$ which are expanded in Fourier basis in the spacetime directions. In our story, after scaling reductions, we can see the insertion of a 3-point interaction vertex as being inserted arbitrarily close to the sphere $\mathbb{CP}^1_{[\mu\bar\lambda]=0}$. In the split theory, such vertex can be seen as the vertex generated after integrating out the bulk graviton modes in $\mathbb{MT}$, which provides Feynman rules in this effective description of the 2d CFT. On spacetime, we would like to interpret this as bringing the spacetime vertex close to the boundary celestial sphere.
\begin{figure}[h]
    \centering
    \includegraphics[scale=0.7]{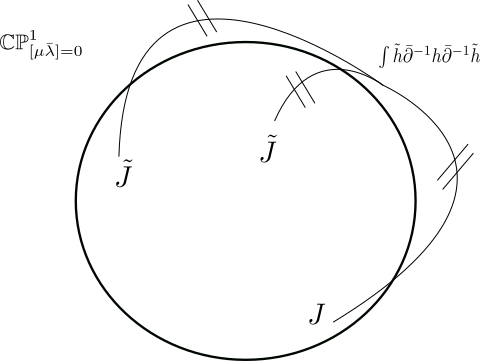}
    \caption{3-point interaction vertex inserted arbitrarily close to $\mathbb{CP}^1_{[\mu\bar\lambda]=0}$.}
    \label{fig:MHV_vertex}
\end{figure}
As shown in figure \ref{fig:MHV_vertex}, the insertion of the 3-point interaction vertex on twistor space restricted to our $\mathbb{CP}^1_{[\mu\bar\lambda]=0}$ enables the existence of additional current correlation functions of the type $\tilde J\tilde J J$. 

In this section, we would like to compute correlation functions of smeared Fourier currents coupling to on-shell graviton scattering states. This is done by providing the 2d CFT with a heavy background vertex sourcing the correlation between three Fourier currents. By adding the following background vertex
\begin{equation}\label{MHV_vertex}
    S_{\text{MHV}}=\int_{\mathbb{CP}^1_{[\mu\bar\lambda]=0}} \D\lambda\wedge \tilde h^{M'}(1)\,\frac{1}{\bar\partial}\,h^{M'}(2)\,\frac{1}{\bar\partial}\,\tilde h^{M'}(3)\frac{\la 13\ra^6}{\la 12\ra\la 23\ra}\,,
\end{equation}
which induces the following non-zero three point function
\begin{multline}
    \la \tilde J(1) J(2)\tilde J(3)\ra_{S_{\text{split}}+S_{\text{MHV}}} =\\
    \left\la\frac{\delta}{\delta \tilde h^{M'}(1)}\,\frac{\delta}{\delta h^{M'}(2)}\,\frac{\delta}{\delta \tilde h^{M'}(3)}\,\e^{\int\D\lambda\wedge\tilde h^{M'}\bar\partial^{-1}h^{M'}\bar\partial^{-1}\tilde h^{M'}\frac{\la 13\ra^6}{\la 12\ra\la 23\ra}}\right\ra_{S_{\text{split}}} = \frac{\la 13\ra^8}{\la 12\ra^2\la 23\ra^2\la 31\ra^2}\,,
\end{multline}
where we have traded the Fourier currents for functional derivatives of graviton Fourier modes $h^{M'}$ and $\tilde h^{M'}$ with the coupling term present in the 2d part of the split action $S_{\text{split}}$ \eqref{eq:split_action}. The insertion of the background vertex was inserted as some effective vertex acted on by the functional derivatives, observe that the $\mathbb{MT}$ part of the action naturally decouples, allowing us to extract the form factor present in the effective vertex. This is precisely the three-point MHV form factor of Einstein gravity.

With the OPE between full currents \eqref{eq:JJ_OPE_full}, we are free to make choices for the reference spinors. For simplicity of the computation, we pick them to be the position of the two negative helicity gravitons on the celestial sphere:
\begin{equation}
    \begin{split}    J(\lambda_i,p_i^{\alpha\dal})J(\lambda_j,p_j^{\alpha\dal})\sim \frac{\D\lambda_j\,[\bar\lambda_i\bar\lambda_j]}{\la ij\ra\la\lambda_i\hat{\lambda}_i\ra\la\lambda_j\hat{\lambda}_j\ra}\frac{\la\lambda_i\lambda_1\ra\la\lambda_i\lambda_3\ra}{\la\lambda_j\lambda_1\ra\la\lambda_j\lambda_3\ra}\,J(\lambda_i,p_i^{\alpha\dal}+p_j^{\alpha\dal}) +i\leftrightarrow j\,.
\end{split}
\end{equation}
It's straightforward to see that this simplifies the computation of higher point correlator by setting all $\tilde J$-$J$ OPEs to $0$. Using this, we can compute four point correlator of the form:
\begin{multline}
    \left\la \tilde J(1)\tilde J(2) J(3)J(4) \right\ra = \left\la \tilde J(1)\tilde J(2) J(3)\right\ra \frac{[34]}{\la 34\ra}\,\frac{\la 13\ra\la 23\ra}{\la 14\ra\la 24\ra}+\left\la \tilde J(1)\tilde J(2) J(4)\right\ra \frac{[34]}{\la 34\ra}\,\frac{\la 14\ra\la 24\ra}{\la 13\ra\la 23\ra}\\
    =\frac{\la 12\ra^6[34]}{\la34\ra\la13\ra\la14\ra\la23\ra\la24\ra}\,,
\end{multline}
which is one of the ways to write the graviton 4-point form factor. Notice the only physical requirement here is that the only non-zero correlator is the three point one sourced by the 2d vertex insertion. It is equivalent to the other ways of writing on the support of momentum conserving delta function. One could check that this OPE procedure would recover all higher multiplicity form factors for gravity scattering \cite{Hodges:2012ym}\footnote{Although one needs to be care when taking OPEs in the correlation functions, one is required to take all possible contractions between the free-fields, where the resulting object does not necessarily recombine into a bilinear current.}. 

\section{Discussion}
To conclude, we have made an attempt to introduce a systematic way of considering celestial holography in 4d asymptotically flat spacetimes. The incorporation of singular conformal primaries wavefunctions using large pure diffeomorphism profiles provides extra field configurations which did not exist in the original theory defined on Fourier-decomposable field configurations. By our "splitting" procedure, we promoted the generators to independent quantum fields on a special locus $[\mu\bar\lambda]=0$ on $\mathbb{PT}^\cO$. This resulted in the formal splitting of the twistor action for self-dual Einstein gravity and we managed to further isolate the dynamics of these free fields by $\mathbb{C}^*$-scaling reduction (Mellin transform in a geometric way). Further investigations of the 2d CFT we derived gives natural coupling between bulk graviton modes and combinations of the free data. 

Although we have obtained (a generalisation in mode numbers of) the $\cL w_{1+\infty}$ algebra from currents coupling to graviton modes localised to different order neighborhoods of the celestial sphere and the leading soft OPE when examining the currents collectively as coupling to graviton scattering states smeared all over $\scri$. It is rather curious how the essentially free large pure diffeomorphism profiles had any access to such detailed information about bulk graviton scattering. We summarize the discussion on the production of form factors in the appendix \ref{Appendix:C}. The key observation here is that, the large pure diffeomorphism profiles moves us in the space of allowed boundary conditions. The fact that our bulk path integral responds to the addition of arbitrarily singular field configurations is this boundary 2d CFT. It sources the addition of these conformal primary wavefunctions to the bulk, which allows for an interpretation of some boundary current. In this sense, our formalism is morally reminiscent of the on-shell soft effective action derivation using edge/large pure diffeomorphism modes \cite{He:2024ddb,Kapec:2021eug}, although the aforementioned papers focused on QED.

It is conceivable to apply the same principles to other 4d theories with uplifts to projective null cone bundle over certain asymptotically flat space due to the nature of the boost eigenstates. In particular, Einstein gravity/conformal gravity and their finite quotients \cite{LeBrun1991ExplicitSM}, even cases including non-zero cosmological constants. We simply follow the procedure of building pure diffeomorphism transformations which are singular at certain designated locus, which sources the splitting on the level of the twistor action once one includes them in the space of field configurations. A further $\mathbb{C}^*$-scaling reduction (geometric Mellin transform) allows us to arrive at the 2d CFT in these theories. 

Unlike those examples in open/closed dualities, the couplings of the two theories are in fact the \textit{same}. However, given the nature of the derivation and the holographic interpretations of the large diffeomorphism profiles, it is quite distinct from a simple rewrite of the theory or the existing worldsheet models that reproduce bulk Einstein gravity \cite{Adamo:2014yya,Adamo:2021bej,Geyer:2014fka}. It certainly is of interest to further unmask the mystery. 

One can further pose the question as in the expectation in usual holography, where bulk graviton modes couple to boundary stress tensor. A natural guess for the stress tensor candidate in our 2d CFT would be to enhance the form degree of the currents $J\in\Omega^{1,0}$ by splitting some trivial integration by parts, which results in (schematically)
\begin{equation}
    \int_{\mathbb{CP}^1_{[\mu\bar\lambda]=0}}\underbrace{\partial^{-1} h}_{\Omega^{-1,1}} \underbrace{\partial J}_{\Omega^{2,0}} + \underbrace{\partial^{-1} \tilde h}_{\Omega^{-1,1}} \underbrace{\partial J}_{\Omega^{2,0}}
\end{equation}
which leaves us with $\partial^{-1} h\in\Omega^{-1,1}$, or it takes values in the holomorphic tangent vector bundle and anti-holomorphic 1-forms. This is precisely the Beltrami differential. The action functional suggests the coupling of the change in complex structure and some stress tensor type object $\partial \tilde J,\,\partial J\in\Omega^{2,0}$. Its OPE with itself clearly has the desired quadratic and double pole structures. 

Regarding the interaction vertices, it would be interesting to understand them as the result of integrating out heavy auxiliary boson and fermion systems as in \cite{Bu:2023vjt,Caron-Huot:2023wdh}, which allows for a straightforward generalisation to vertices generating N$^k$MHV degree form factors in gravity. Another puzzle we currently have is the background deformation operator we are permitted to add. There seems to be no generic restriction or first principle derivation of the effective vertices we have. We would like to study this further in future work.

\acknowledgments
It is a pleasure to thank Tim Adamo and David Skinner for interesting discussions. We also thank Sonja Klisch for proofreading the manuscript. WB is supported by the Royal Society Studentship. SS is supported by the Trinity Internal Graduate Studentship. The work of SS has been supported in part by STFC HEP Theory Consolidated grant ST/T000694/1.

\newpage
\appendix
\section{Uplifting free scalar theory to twistor space}\label{appendix_A}
As a warmup, consider the uplift of the free scalar theory on $\mathbb{R}^{4-p,p}$ to a theory of 1-forms on $\mathbb{R}^{4-p,p}\times \mathbb{CP}^1$ (for $p=0,1,2$):
\begin{equation}
    \int_{\mathbb{R}^{4-p,p}} \, \d^4x \, \Phi\square\Phi \rightarrow \int_{\mathbb{R}^{4-p,p}} \Omega \wedge \Phi \d \Phi\,.
\end{equation}
The case of $p=0$ is familiar to us as the projective twistor space $\mathbb{PT}$, a complex 3-fold. In that case, the space of 1-forms admits a decomposition into the 3 dimensional space of $(0,1)$ forms and the 3 dimensional space of $(1,0)$ forms, related by Euclidean conjugation.

\begin{equation}
    \Omega^{1,0}:=\{\D \lambda := \langle \lambda \d \lambda \rangle, e^{\dal} :=\d(\langle x^{\dal} \lambda\rangle) \}, \quad \Omega^{0,1}:=\left\{\bar e^0 :=\frac{\D \hat\lambda}{\langle \lambda \hat \lambda \rangle^2}, \bar e^{\dal}:=\d(\langle x^{\dal} \hat\lambda\rangle) \right\}\,.
\end{equation}
Their dual vectors are
\begin{equation}
    T^{1,0}:=\left\{ \frac{1}{\langle \lambda \hat \lambda \rangle}\left\langle \hat \lambda\frac{\partial}{\partial \lambda}\right\rangle, \frac{1}{\langle \lambda \hat \lambda \rangle} \left\langle \hat \lambda\frac{\partial}{\partial x^{\dal}}\right\rangle\right\}\,;
\end{equation}
\begin{equation}
    T^{0,1}:=\left\{ \langle \lambda \hat \lambda \rangle \left\langle \lambda \frac{\partial}{\partial \hat \lambda}\right\rangle, \left\langle \lambda \frac{\partial}{\partial x^{\dal}}\right\rangle\right\}\,.
\end{equation}
For the case $p\neq 0$ we can still go ahead and define the 3 dimensional space of "$(0,1)$" forms and "$(1,0)$" forms, related to the $p=0$ forms by analytic continuation in the $x^{\alpha \dot \alpha}$ coordinate. The $(0,1)$ vector fields (resp. $(1,0)$) still form an involutional distribution. However, the $(0,1)$ and $(1,0)$ forms and vectors are no longer related by complex conjugation. For any value of $p$, we define $\Omega$ to be the following $\d$-closed $(3,0)$ form:
\begin{equation}
    \Omega:=\D \lambda \wedge [\d(\langle x \lambda\rangle)\d(\langle x \lambda\rangle)]\,.
\end{equation}
In the theory defined on $\mathbb{R}^{4-p,p}\times \mathbb{CP}^1$, the fields are not scalars but 1-forms twisted by $\mathcal{O}(-2)$ (i.e they are homogeneous of degree -2 as functions of $\lambda^{\alpha}$). Due to the presence of the wedged $\Omega$ in the action, the action only depends on the $(0,1)$ components of $\phi$, and the $(0,1)$ component of $\d$ that we conventionally name $\bar \partial$. We name the relevant components of $\bPhi$ as
\begin{equation}
    \bPhi = \bPhi_0 \bar e^0 + \bPhi_{\dal} \bar e^{\dal}\,.
\end{equation}
The theory on $\mathbb{R}^{4-p,p}\times \mathbb{CP}^1$ has a cohomological symmetry due to the nilpotence of $\d$:
\begin{equation}
    \bPhi \rightarrow \bPhi + \bar \partial \Lambda, \quad \Lambda \in \Omega^{0}(\mathbb{R}^{4-p,p}\times \mathbb{CP}^1, \mathcal{O}(-2))\,.
\end{equation}
In order to see the classical equivalence of the actions, the first step is to use the cohomological symmetry to fix "spacetime gauge", in which $\bPhi$ restricted to the $\mathbb{CP}^1$ is harmonic with respect to the usual round metric on the $\mathbb{CP}^1$. Together with the fact that $\bPhi_0$ is a smooth section of $\mathcal{O}$, we find that $\bPhi_0$ must be independent of $\lambda^{\alpha}$:
\begin{equation}
    \partial \bPhi_0 = 0 \implies \bPhi_0 = \Phi(x)\,.
\end{equation}
We can integrate out $\bPhi_{\dot \alpha}$ by completing the square:
\begin{equation}
    S = \int \Omega \wedge \bar \Omega \left(\bPhi^{\dot \alpha}(\bar \partial_0 \bPhi_{\dot \alpha}-2 \bar \partial_{\dot \alpha}\bPhi_0)\right) \rightarrow \int \frac{\D \lambda \D \hat \lambda}{\langle \lambda \hat \lambda \rangle^2}\, \d^4 x \, \left(- \bar \partial^{\dot \alpha}\bPhi_0 \bar \partial^{-1}_0\bar \partial_{\dot \alpha}\bPhi_0)\right)\,.
\end{equation}
By inspection, $\bar \partial^{-1}_0 \bar \partial_{\dot \alpha} \bPhi_0 = \partial_{\dot \alpha} \bPhi_0$ and therefore we have:
\begin{equation}
    S = -\int \frac{\D \lambda \D \hat \lambda}{\langle \lambda \hat \lambda \rangle^2}\, \d^4 x \, \Phi(x) \square \Phi(x) = -4 \pi \int \d^4 x \, \Phi(x) \square \Phi(x)\,.
\end{equation}
Notice that the derivation was totally agnostic of the signature of the spacetime $\mathbb{R}^{4-p,p}$. We are therefore sometimes somewhat sloppy about the signature of the underlying spacetime, as the analytic continuation to change $p$ commutes with all the standard twistorial manipulations we will use.

\section{Lorentzian story}\label{appendix:Lorentzian_story}
\subsection{The quotient space $\mathcal{M^{O}}/\{x \sim s x, s \in \mathbb{R}^+\}$}\label{subsec:quotient_space}
We have laid the groundwork for decomposing 4d quantum fields into Mellin modes, integrating out the radial direction, and working directly on $\mathcal{M^{O}}/\{x \sim s x, s \in \mathbb{R}^+\}$. The advantage of this approach is that the theory now lives in 3 spacetime dimensions, and celestial holography can be interpreted as a standard codimension one holography, possibly amenable to standard holographic techniques. In this section we will discuss how to think about the quotient space $\mathcal{M^{O}}/\{x \sim s x, s \in \mathbb{R}^+\}$. For more detail on the geometric set up of the problem, the readers are referred to our previous papers \cite{Bu:2023cef,Bu:2023vjt}.

The space $\mathcal{M^{O}}/\{x \sim s x, s \in \mathbb{R}^+\}$ is most conveniently described in terms of 4 coordinate patches. First, there is the null cone of the origin, $\{x^2 = 0\}$. The other three patches are each connected component of the remaining space, $\mathcal{M^O}\setminus \{x^2=0\}$. First, there are the two Milne regions, being the locus $\{t>0, x^2 < 0\}$ and the locus $\{t<0, x^2 < 0\}$, where $t=x^0$ is the time direction. Then, there is the Rindler region, $\{x^2 > 0\}$. Although there are many other ways to pick coordinate patches of the quotient space $\mathcal{M^{O}}/\{x \sim s x, s \in \mathbb{R}^+\}$, the partition into Milne, Rindler, and lightcone is (proper orthochronous) Lorentz invariant and therefore the most convenient for our purposes.

On each of the Milne and Rindler patches, we have the 3 independent local coordinates:
\begin{equation}
   P_{\text{M+R}}:\quad \frac{x^{\alpha \dot \alpha}}{\sqrt{|x^2|}}, \quad x^{\alpha \dot \alpha} \in \frac{\{x^2\neq 0\}\subset \mathcal{M^O}}{x \sim s x, s \in \mathbb{R}^+}
\end{equation}
and we can define the invariant and basic metric (the Lie derivative under the Euler vector field $\Upsilon =x \cdot \frac{\partial}{\partial x}$ vanishes and contraction with the Euler vector field vanishes):
\begin{equation}
    \d s_{3d}^2 := \frac{1}{x^2}\left(\d x^{\mu} - \frac{x^\mu x \cdot \d x}{x^2}\right)^2 = \frac{\d x^2}{x^2} - \frac{(x \cdot \d x)^2}{x^4}
\end{equation}
On each Milne patch, this metric is the 3d Euclidean Anti-de Sitter, or EAdS$_3 =\mathbb{H}_3$ metric. On the Rindler patch, this metric is the 3d Lorentzian de Sitter (LdS$_3$) metric.
On the null cone of the origin, this metric is unsuitable because it is singular. As is familiar from AdS$_3$/CFT$_2$, under a conformal rescaling of the 3d line element $\d s^2_{3d} \rightarrow x^2 \d s^2_{3d} $, we can recover the round metric on the two $S^2$s that are the shared conformal boundary of H$^3$ and the two timelike infinities of LdS$_3$. We have two independent local coordinates
\begin{equation}
    P_{\text{cone}}:\quad\frac{x^{\alpha \dot \alpha}}{t}, \quad x^{\alpha \dot \alpha} \in \frac{\{x^2=0\}\subset \mathcal{M^O}}{x \sim s x, s \in \mathbb{R}^+}
\end{equation}
These local coordinates are valid on neighborhoods that extend off of $\{x^2=0\}$, as long as they do not contain $t=0$. The transition function to go to the other patch on such neighborhoods is of course multiplication by $t/\sqrt{|x^2|}$.

Therefore, the way to think about the quotient space $\mathcal{M^{O}}/\{x \sim s x, s \in \mathbb{R}^+\}$ is as a copy of LdS$_3$ with two $\mathbb{H}_3$ caps glued onto the $S^2$s at positive and negative timelike infinity. From a field $\Phi$ on $\mathcal{M^O}$, a scaling decomposition results in infinitely many scaling invariant fields $\Phi_{\Delta}$ living on the quotient space $\mathcal{M^{O}}/\{x \sim s x, s \in \mathbb{R}^+\}$, indexed by $\Delta\in\mathbb{R}$. Each $\Phi_{\Delta}$ is the coefficient of a scaling weight basis function $Y_{\Delta}$ that was an eigenstate of scaling:
\begin{equation}
    \Phi = \int \d \Delta \Phi_{\Delta} Y_{\Delta}, \quad (x \cdot \partial_x)\Phi_{\Delta}=0, \quad  (x \cdot \partial_x)Y_{\Delta} = \Delta Y_{\Delta}
\end{equation}
More formally, we can interpret $\mathcal{M^O}$ as the total space of an $\mathbb{R}^+$ bundle over the quotient space. Covering the base space with open sets:
\begin{equation}
    \{x^2<0\}, \{t \neq 0\}, \{x^2 > 0\}
\end{equation}
With respect to the local coordinates defined above, we define an $\mathcal{O}(n)$ bundle as the line bundle with transition functions:
\begin{align}
    f(\{x^2>0\},\{t \neq 0\}) = \left(\frac{\sqrt{x^2}}{t}\right)^n, \nonumber \\
    f(\{x^2<0\},\{t \neq 0\}) = \left(\frac{t}{\sqrt{-x^2}}\right)^n
\end{align}
As a function on the total space $\mathcal{M^O}$, $\Phi_{\Delta}Y_{\Delta}$ is a scaling eigenstate with eigenvalue $\Delta$. It is to be interpreted as a section of $\mathcal{O}(\Delta)$ over the base space $\mathcal{M^O}/\mathbb{R}^+$. $\Phi_{\Delta}Y_{\Delta}$ can then be written in local coordinates as a field on each $\mathbb{H}_3$, LdS$_3$, or $S^2$ in a perfectly finite way. The transition function between the Rindler/Milne region and the null cone can be described by the usual limiting procedure
\begin{align}
     \left(\lim_{x^2 \rightarrow 0} \Phi_{\Delta}(x) \left(\frac{\sqrt{\pm x^2}}{t}\right)^{\Delta}\right)_{P_{\text{M+R}}} = (\Phi_{\Delta}(x))_{P_{\text{cone}}}
\end{align}
We recognise multiplying in the transition function as the familiar procedure to recover the rescaled boundary value of a bulk field of conformal weight $\Delta$ in the embedding space description of conventional AdS$_3$/CFT$_2$.



\subsection{$\mathbb{C}^*$ scaling reduction on twistor space}\label{subsec:scaling_reduction}

We have set the stage to decompose our quantum fields into eigenstates of scalings. Having done so, we can integrate out the radial spacetime direction and descend to a theory with 3 spacetime directions. Consider decomposing fields $\bPhi(x,\lambda)$ on $\mathcal{M^O}\times \mathbb{CP}^1$ into scaling modes. Working on the patches where $\{x^2 \neq 0\}$ (we will do each patch separately and patch them together with the above transition functions),
\begin{equation}
    \bPhi(x,\lambda) = \int \d \Delta \bPhi_{\Delta}(x, \lambda) \left(\sqrt{|x^2|}\right)^{\Delta}, \quad (x \cdot \partial_x)\bPhi_{\Delta}(x, \lambda)=0
\end{equation}
\begin{equation}
    \bPhi(x,\lambda), \, (x,\lambda) \in \mathcal{M^{O}}\times \mathbb{CP}^1 \xrightarrow{\text{scaling decomp.}} \bPhi_{\Delta}(x,\lambda), \, x \in \mathcal{M^{O}}/\{x \sim s x, s \in \mathbb{R}^+\}\times \mathbb{CP}^1, \Delta \in \mathbb{R}
\end{equation}
Since this space is odd-dimensional, it cannot be given a complex structure, which is inconvenient for our purposes. We would prefer to perform another mode decomposition and integrate out another spacetime dimension such that the spacetime we work on is even dimensional. The traditional choice in twistor dimensional reduction is to integrate out motion along geodesics, which we implement in the following way. First consider distinct real Lorentzian null momenta defined up to scale, $k_1^{\alpha \dot \alpha}\sim\lambda^{\alpha}\bar \lambda^{\dot \alpha}, k_2^{\alpha \dot \alpha} \sim \nu^{\dot\alpha}\bar\nu^{\alpha}$.
\begin{equation}
    (k_1, k_2) \in \{ S^2 \times S^2 \}\setminus S^2_{k_1 \propto k_2}
\end{equation}
In $\mathcal{M^{O}}/\{x \sim s x, s \in \mathbb{R}^+\}$, the ordered pair $(k_1,k_2)$ defines a piecewise geodesic curve in the following way. In the Milne regions of $\mathcal{M^O}/\{x \sim s x, s \in \mathbb{R}^+\}$, $(k_1,k_2)$ define the endpoints of an oriented spacelike geodesic of a H$^3$ in the future of the origin and another oriented spacelike geodesic of a $\mathbb{H}_3$ in the past of the origin. In the Rindler region of $\mathcal{M^O}/\{x \sim s x, s \in \mathbb{R}^+\}$, $(k_1,k_2),(k_2,k_1)$ define the endpoints of a pair of timelike geodesics in the LdS$_3$, one which starts at $k_1$ in the infinite timelike past and ends at $k_2$, and vice versa. Fields $\bPhi_{\Delta}$ living on these piecewise geodesic curves can now be decomposed into modes along the curve. Since each component of the curve is non-compact, the mode number for this decomposition is another continuous parameter.
\begin{equation}
    \bPhi_{\Delta}(x,\lambda), \, x \in \mathcal{M^{O}}/\mathbb{R}^+, \lambda \in \mathbb{CP}^1, \Delta \in \mathbb{R} \xrightarrow{\text{geodesic motion}} \bPhi_{\vec \Delta}(k_1,k_2), k_1,k_2 \in S^2\times S^2 \setminus S^2, \vec \Delta \in \mathbb{R}^2
\end{equation}
We recognise the final spacetime as $\mathbb{MT}$, the minitwistor space of $\mathbb{H}_3$. In this way, we show that a theory of fields $\bPhi_{\vec \Delta}$ on $\mathbb{MT}$ carrying two continuous mode numbers is the mode decomposition of a theory of fields $\bPhi$ on the $\{x^2 \neq 0\}$ patch of $\mathcal{M^O}\times \mathbb{CP}_{\lambda}^1$. Notice that it is crucial that $k_1$ was distinct from $k_2$ as a projective null momentum. If they become collinear, each geodesic in the two $\mathbb{H}_3$ factors degenerates into a single boundary point, and the pair of timelike LdS$3$ geodesics coincide, go null, and leave the LdS$_3$ hypersurface. This picks out a single point $k_1$ on the celestial sphere $\{x^2=0\}$. Of course, this is precisely the antidiagonal $\mathbb{CP}^1$ that is removed when we define $\mathbb{MT}$ as the direct product of $\mathbb{CP}^1_{\mu} \times \mathbb{CP}^1_{\lambda} \setminus \{[\mu \bar \lambda] = 0\}$. The patch where $\{x^2=0\} \subset \mathcal{M^O}$ is therefore simply the removed $\mathbb{CP}^1$, and adding it back in gives us the $\mathbb{C^*}$ dimensionally reduced theory on $\mathcal{M^O}\times \mathbb{CP}^1$.

\section{MHV form factors from current correlators on twistor space}\label{Appendix:C}

We will discuss a quick and dirty way of getting the Fourier currents that is unfortunately conceptually a little winding, which is why it is excluded from the main text. Computationally, it is quite efficient, and we explain it here.

As a quick refresher, in section \ref{subsec:2d_CFT_currents} where we discuss the 2d CFT, recall that the basis decomposition in the main text decomposes fields into Laurent series in $[\mu \bar \lambda]=0$ and selects out the part of the action that is independent of $[\mu \bar \lambda]$. 
\begin{align}
    &\int \bar \delta^{(k-1)}([\mu \bar \lambda]) \tilde \phi_k \frac{(-1)^k}{k!}\prod_i \mathcal{O}_i(Z^A, \bar Z^A) = \int \bar \delta^{(k-1)}([\mu \bar \lambda]) \tilde \phi_k \frac{(-1)^k}{k!} \prod_i \left(\sum_{p_i} (\mathcal{O}_i)_{p_i} \left(\frac{[\mu \bar \lambda]}{\la \lambda \hat \lambda \ra}\right)^{p_i}\right)\nonumber
    \\
    &=\int_{{[\mu \bar \lambda]=0}\subset \mathbb{PT}^\cO} \tilde \phi_k \delta\left(\sum p_i - k\right) \prod_i \left(\sum_{p_i} (\mathcal{O}_i)_{p_i}  \right)\nonumber\,.
\end{align}
The remaining component of $\mu$ is the $[\mu \hat{\bar\lambda}] \neq 0$, which is a $\mathbb{C}^*$ in local coordinates. Using a Fourier series in the angular parameter and a Fourier transform in $\log$ of the radial parameter, we integrate out the remaining component of $\mu$. We therefore land on the 2d CFT on $\{[\mu \bar \lambda]=0\}/\{\mu^{\dal} \sim s \mu^{\dal}, s \in \mathbb{C}^*\}$

This procedure can be reversed to reconstruct effective $\mathbb{PT}^{\mathcal{O}}$ fields by summing over basis functions for each decomposition. It is important to stress, however that this is not the same as the original theory. Major differences are:
\begin{enumerate}
    \item Tilded fields were not present in the unsplit $\mathbb{PT}^{\mathcal{O}}$ theory to begin with
    \item Untilded fields are not fully reconstructed! Rather, their components that had a Laurent expansion (and therefore were sampled by the introduction of the $\bar \delta^{(k-1)}([\mu \bar \lambda])\tilde \phi$ (resp. $\tilde \eta$) and localise to $[\mu \bar \lambda]=0$) are collected and summed into an effective $\mathbb{PT}^{\mathcal{O}}$ field
    \item The dependence of fields on $[\mu \hat{\bar \lambda}]$ and its complex conjugate is reconstructed faithfully by the inverse Fourier transform. However, it is not so for the $\la \bar \mu \lambda \ra$. Neither tilded nor untilded effective fields depend on $\la \bar \mu \lambda \ra$! Tilded fields never had any dependence to begin with, and for untilded fields we lost all information of $\la \bar \mu \lambda \ra$ once we are on $\{[\mu \bar \lambda]=0\}$. Resumming the modes, we only reinsert $[\mu \bar \lambda]$ dependence:
    \begin{equation}
        \mathcal{O}_{\text{eff}}:= \sum_{p} \mathcal{O}_{p}([\mu \hat{\bar \lambda}], \lambda^{\alpha}, c.c) \left(\frac{[\mu \bar \lambda]}{\la \lambda \hat \lambda \ra}\right)^{p}
    \end{equation}
    This is a choice made to simplify the form of the resulting effective $\mathbb{PT}^\cO$ theory. It reincorporates the mass term in the 2d theory and does not introduce new terms.
    \item As $(0,1)$-forms, the $\la \lambda \d \bar \mu \ra$ component of untilded fields was never sampled, as the form in that direction was saturated by the presence of the $\bar\delta^{(k-1)}([\mu \bar \lambda])$
\end{enumerate}
Therefore the effective $\mathbb{PT}^\cO$ theory is quite a different beast than the original twistor action. It takes the form:
\begin{align}
    &\int \frac{\D^3 Z}{\text{Vol}(\mathbb{C})} \tilde H \wedge \left(\bar\partial H + \frac{1}{2} \{H,H\}\right)\nonumber
    \\
    &H := (h - h_{\la \lambda \d \bar \mu \ra \text {component}})_{\text{projected and Laurent resummed}} + \sum_k \tilde \eta_k \left(\frac{\la \lambda \hat \lambda \ra}{[\mu \bar \lambda]}\right)^k \frac{\la \lambda \d \bar \mu \ra}{\la \lambda \hat \lambda \ra}\,,
\end{align}
and likewise for $\tilde H$. The $\frac{1}{\text{Vol}(\mathbb{C})}$ factor is in the measure to balance the result of integrals over $z=\frac{[\mu \bar \lambda]}{\la \lambda \hat \lambda \ra}$
\begin{equation}
    \int \d z \d \bar z \,\, z^a = \text{Vol}(\mathbb{C}) \delta(a)\,,
\end{equation}
and can be explicitly implemented with a Gaussian kernel. The awkwardness with the Vol$(\mathbb{C})$ is the price that we have to pay to resum all the fields into these master fields on $\mathbb{PT}^\cO$ that simplify calculations and notations significantly.

Note that the $J_{\Delta, k}$ current has been resummed into all the terms that couple to the $\D \hat \lambda$ component of $H$. Now consider the cubic vertex:
\begin{equation}
    \int \D^3 Z \, H \wedge \{H, \tilde H\}\,.
\end{equation}
It is more convenient to take the Fourier decomposition\footnote{suitably regulated, as of course these fields are not in general square integrable in the $x$ factor} in order to diagonalise with respect to the derivatives.
\begin{equation}
    H(\lambda, \mu) = \int \d^4k \, e^{ik \cdot x} H(\lambda, k_{\alpha \dot \alpha}) = \int \d^4k \, e^{i\left(\frac{[\mu k \hat \lambda\rangle-[\hat \mu k \lambda\rangle}{\langle \lambda \hat \lambda \rangle}\right)} H(\lambda, k_{\alpha \dot \alpha})
\end{equation}
\begin{equation}
    \frac{\partial}{\partial \mu^{\dal}} = \frac{\hat \lambda^{\alpha}}{\la \lambda \hat \lambda \ra}\frac{\partial}{\partial x^{\alpha \dot \alpha}} \quad \implies \quad \{e^{\im k_1\cdot x},e^{\im k_2\cdot x}\} = \frac{\langle \hat \lambda k_1^{\dot \alpha}\rangle\langle \hat \lambda k_{2\dot \alpha}\rangle}{\langle \lambda \hat \lambda \rangle^2} e^{\im (k_1+k_2) \cdot x}\,.
\end{equation}
For future convenience, name the components of $H = H_{\hat \lambda}+H_{\dal}\bar e^{\dal}$, and likewise for $\tilde H$.
\begin{align}
    \int \D^3 Z \,H_{\hat\lambda} \wedge \{H, \tilde H\} = \int\,\d^4k \, \D\lambda \, H_{\hat \lambda}(\lambda, k_{\alpha \dot \alpha}) \int \d^2 \mu \, e^{i\left(\frac{[\mu k \hat \lambda\rangle-[\hat \mu k \lambda\rangle}{\langle \lambda \hat \lambda \rangle}\right)} \{H, \tilde H \}
    \\
    =\int\,\d^4k_1\d^4k_2 \, \D\lambda \, H_{\hat\lambda}(\lambda, -k_1-k_2) \frac{\langle \hat \lambda k_1^{\dot \alpha}\rangle\langle \hat \lambda k_{2\dot \alpha}\rangle}{\langle \lambda \hat \lambda \rangle^2}[H(\lambda, k_1), \tilde H(\lambda, k_2)]\,,
\end{align}
$H_{\dal}\in \Omega^0(\mathcal{O}(3)), \tilde H_{\dal}\in \Omega^0(\mathcal{O}(-5))$ are 0-forms with the following kinetic term and we can read off the propagator.
\begin{equation}
    \int \d^4k\,\D \lambda \wedge [H(\lambda,-k) \, \bar \partial_{\lambda} \tilde H(\lambda,k)], \quad H_{\dal}(\lambda_1,k_1)\tilde H_{\dot \beta}(\lambda_2,k_2)\sim \frac{\epsilon_{\dal \dot \beta}\delta^4(k_1+k_2)}{\langle \lambda_1 \lambda_2 \rangle} \left(\frac{\langle i \lambda_1 \rangle}{\langle i \lambda_2 \rangle}\right)^4\,.
\end{equation}
Now we see that $J \in \Omega^{0,1}(\mathbb{PT}, \mathcal{O}(-2))$, the resummed version of $J_{\Delta,k}$, takes the form:
\begin{equation}
    J(\lambda,k^{\alpha \dot \alpha})=\D \lambda \wedge \int \d^4 p \frac{\langle \hat \lambda k^{\dot \alpha}\rangle\langle \hat \lambda p_{\dot \alpha}\rangle}{\langle \lambda \hat \lambda \rangle^2}[H(\lambda, (k-p)^{\alpha \dot \alpha}), \tilde H(\lambda, p^{\alpha \dot \alpha})]\,.
\end{equation}
A largely identical computation gives the resummed version of $\tilde J_{\Delta,k}$, by reading off the term that couples to $\tilde H_{\hat \lambda}$. This is called $\tilde J(\lambda,k^{\alpha \dot \alpha}) \in \Omega^{0,1}(\mathbb{PT}, \mathcal{O}(6))$:
\begin{equation}
    \tilde J(\lambda,k^{\alpha \dot \alpha})=\D \lambda \wedge \int \d^4 p \frac{\langle \hat \lambda k^{\dot \alpha}\rangle\langle \hat \lambda p_{\dot \alpha}\rangle}{\langle \lambda \hat \lambda \rangle^2}[H(\lambda, (k-p)^{\alpha \dot \alpha}), H(\lambda, p^{\alpha \dot \alpha})]\,.
\end{equation}
In the kinematic prefactor we could write $(k-p)^{\alpha \dot \alpha}$ instead of $p^{\alpha \dot \alpha}$, but the difference between the expressions vanishes due to antisymmetry, so to save space we will write $p^{\alpha \dot \alpha}$. The $JJ$ OPE is therefore:
\begin{multline}
    J(\lambda_1,k_1^{\alpha \dot \alpha})J(\lambda_2,k_2^{\alpha \dot \alpha})\sim \frac{\delta^4(k_1+k_2)}{\langle \lambda_1 \lambda_2 \rangle^2}\left(\frac{\langle \hat \lambda_1 k_1^{\dot \alpha}\rangle\langle \hat \lambda_2 k_{2\dot \alpha}\rangle}{\langle \lambda_1 \hat \lambda_1 \rangle\langle \lambda_2 \hat \lambda_2 \rangle}\right)^2 \\
    + \frac{\D \lambda_1}{\langle \lambda_1 \lambda_2 \rangle}\, \frac{\langle \hat \lambda_1 k_1^{\dot \alpha}\rangle\langle \hat \lambda_2 k_{2\dot \alpha}\rangle}{\langle \lambda_1 \hat \lambda_1 \rangle\langle \lambda_2 \hat \lambda_2 \rangle}\left(\frac{\langle i \lambda_2 \rangle}{\langle i \lambda_1 \rangle}\right)^2J(\lambda_2,(k_1+k_2)^{\alpha \dot \alpha})\,.
\end{multline}
These can be interpreted in terms of Feynman diagrammatics in the effective $\mathbb{PT}^\cO$ theory. The double pole is a 1-loop effect (the $H$ self-energy). If we compute correlators with these OPE rules, we get rational functions of $\lambda_j, k_j$. If we analytically continue $k_{j}$ to null real Lorentzian momenta built from $\lambda, \bar \lambda$, we are permitted to choose:
\begin{equation}
    k_{\alpha \dot \alpha}(\lambda):=\omega \frac{\lambda_{\alpha}\bar\lambda_{\dot \alpha}}{\langle \lambda \hat \lambda \rangle}\,.
\end{equation}
Setting $k$ to this value for each $J$, the simple pole in the $J(\lambda_1,k_1^{\alpha \dot \alpha}(\lambda_1))J(\lambda_2,k_2^{\alpha \dot \alpha}(\lambda_2))$ OPE becomes:
\begin{equation}
     \frac{\D \lambda_1}{\langle \lambda_1 \lambda_2 \rangle}\, \frac{[\bar \lambda_1 \bar \lambda_2]}{\langle \lambda_1 \hat \lambda_1 \rangle\langle \lambda_2 \hat \lambda_2 \rangle}\left(\frac{\langle i \lambda_2 \rangle}{\langle i \lambda_1 \rangle}\right)^2J(\lambda_2,k_1(\lambda_1)+k_2(\lambda_2))\,.
\end{equation}
The simple pole is the splitting function for $++\rightarrow +$ gravitons. Now consider MHV-type correlators made of insertions of
\begin{equation}
    \left \la \tilde J \tilde J J J ... J \e^{S_{\text{MHV}}}\right \ra\,.
\end{equation}
It has all the right singularities to be the MHV form factor, and of course the collinear limits are precisely the gravity splitting functions. After some work, it can be seen that the computation of these correlators in fact gives the Tree diagram formalism for MHV amplitudes \cite{Bern:1998sv,Nguyen:2009jk}, which is equivalent to the Hodges formula \cite{Hodges:2012ym}.

\section{Charges for diffeomorphisms}\label{sec:charges} 
Consider a gauge theory with Lagrangian $L$, defined on a manifold with a nonempty set of (ordinary or conformal) boundaries or defects $\partial M$. In order to specify the theory, we must specify boundary conditions $B$ on $\partial M$ in order to say exactly what sorts of field configurations we wish to sum over. These boundary conditions often break some gauge invariance. For example, a Dirichlet boundary condition on $U \subset \partial M$ for a field subject to a gauge transformation breaks all gauge transformations that do not vanish at $U$. Hence in order to incorporate path integral over the correct configurations, one needs compatible boundary conditions and the allowed gauge transformations. In particular, large gauge transformations refers to gauge transformations which do not fall off at $\partial M$, it represents residual degree of freedom of the bulk theory which allows further changes in the configuration of the fields with certain behaviour at $\partial M$. For compatibility reasons, such gauge transformations modify the boundary conditions to include previously not allowed field configurations. In the case of twistor space $\mathbb{PT}^\cO$ of 4d Minkowski space, we recognize $\partial M$ as the extra sphere that appears in the split theory $\mathbb{CP}^1_{[\mu\bar\lambda]=0}$. The set of large gauge transformations we are interested in in this paper are diffeomorphisms of vacuum Einstein equation in 4d that do not vanish at infinity, these correspond to radiation degrees of freedom which further permits certain field configurations in the 4d bulk. On twistor space, after scaling reduction and mode decomposition in geodesics, instead of fixing the set of boundary conditions on the special defect sphere $\mathbb{CP}^1_{[\mu\bar\lambda]=0}$, we promote the boundary conditions to a dynamical field and integrate over them. The fields here are precisely the radiative degrees of freedom living on the celestial sphere uplifted to the twistor sphere. Therefore, we can construct large gauge charges that act on the boundary configurations and the bulk fields at the same time.

In this section, we shall discuss the asymptotic symplectic charge induced by large pure diffeomorphism transformations which localise on the locus $u=0$. Recall that such shifts in the bulk gravitons 
\begin{equation}
    h\to h+\bar\partial h_G
\end{equation}
amounts to shifting the combination $h'+\bar\delta([\mu\bar\lambda]=0)h_G$ in the "split" theory, for the redefined bulk field $h'$ and the boundary profile $h_G$:
\begin{equation}
    h'\to h+\bar\delta([\mu\bar\lambda]=0) f \,,\quad h_G\to h_G- f
\end{equation}
We begin with Einstein gravity in the self-dual sector on twistor space with the origin removed $\mathbb{PT}^\cO$
\begin{equation}
    S_{\text{SD}}=\int_{\mathbb{PT}^\cO} \D^3Z\wedge \tilde h\wedge \left(\bar\partial h+\frac{\partial}{\partial\mu^{\dal}}h\,\frac{\partial}{\partial\mu_{\dal}}h\right)
\end{equation}
The symplectic potential on the phase space can be obtained from this by varying the action twice and use the fact that $\delta^2 S_{\text{SD}}=0$:
\begin{equation}
    \int_{\mathbb{PN}} \D^3Z\wedge \delta h\wedge \delta \tilde h
\end{equation}
Regardless of the signature of spacetime we are considering here, twistor space would fibre over the 5d real manifold $\mathbb{PN}$ under the CR condition 
\begin{equation}
    \la\bar\mu\lambda\ra+[\mu\bar\lambda]=0
\end{equation}
On $\mathbb{PN}\cong\mathbb{R}^3\times S^2$, the $\mu$ coordinates we have are 
\begin{equation}
    \underbrace{[\mu\bar\lambda],\quad [\hat{\mu}\bar\lambda],\quad [\mu\hat{\bar\lambda}]}_{\mathbb{PN}\,\text{coordinates}},\quad \underbrace{[\hat{\mu}\hat{\bar\lambda}]=\la\bar\mu\lambda\ra}_{=-[\mu\bar\lambda]\,\text{on the CR condition}}
\end{equation}
To get to $\mathbb{C}_u\times S^2$, we just need to reduce on $q=[\mu\hat{\bar\lambda}]$, which is the $\mathbb{R}^3$ minitwistor space direction. In order to obtain the charge for the boundary values of the bulk fields, we begin by writing the symplectic potential in the components
\begin{equation}
    \int_{\mathbb{PN}}\D^3Z\wedge \underbrace{\tilde h}_{\lambda\,\text{direction}}\wedge \bar\partial_\mu h_G +\underbrace{\tilde h}_{\mu\,\text{direction}}\wedge \bar\partial_\lambda h_G
\end{equation}
where $\tilde h$ is an antiholomorphic 1-form pointing in the $\lambda$ direction, $h_G$ is the large gauge profile we have earlier, the antiholomorphic $\mu$ derivative serves as a localisation differential, we shall see that it picks up poles of $h_G$ and localises the integral to some order formal neighborhood of the celestial sphere. Note that the second drops out since we make explicit gauge choice such that the asymptotic News tensor does not have a $u$-component, hence $\tilde h_{\mu}=0$. We only keep the first term, which can be further evaluated:
\begin{equation}
    \int_{5d} \frac{\d u\wedge\D\lambda}{\la\lambda\hat{\lambda}\ra}\wedge \d q \wedge \tilde h\wedge\left(\frac{[\bar\lambda\d\hat{\mu}][\hat{\bar\lambda}\hat{\mu}][\mu\bar\lambda]}{[\mu\hat{\mu}]\la\lambda\hat{\lambda}\ra}\,\frac{\partial}{\partial[\hat{\mu}\bar\lambda]}+\frac{[\hat{\bar\lambda}\d\hat{\mu}][\mu\bar\lambda][\bar\lambda\hat{\mu}]}{[\mu\hat{\mu}]\la\lambda\hat{\lambda}\ra}\frac{\partial}{\partial[\hat{\mu}\bar\lambda]} \right)h_G
\end{equation}
where we have Schoutened $\bar\partial_\mu = \frac{[\hat{\mu}\d\hat{\mu}]\mu^{\dal}}{[\mu\hat{\mu}]}\frac{\partial}{\partial \hat{\mu}^{\dal}}$ on the numerator, the second term in the bracket vanishes when acting on our $h_G$, since it does has weight $0$ in $[\hat{\mu}\bar\lambda]$. We have picked the pure diffeo profile $h_G=q^{k+1}\la\lambda\hat{\lambda}\ra^k\,\tilde\eta_{-2k}/u^k$, for $k\in\mathbb{Z}_+$. Parametrizing the $q$-integral with polar coordinate with some fixed radius $r$, we shall take the radius to infinity towards the end. In the meantime, we also pick the boundary mode $\tilde h=h_{\bar\lambda}\,f(q)$ to have some pure boundary radiative degree of freedom $h_{\bar\lambda}$ and certain $q$-dependent function $f(q)$. The $u$-component of the boundary radiative mode vanishes, so this is practically our only way to distribute the form degrees. The integral can be written as 
\begin{equation}
    \int_{0}^{2\pi} \im\d\theta(r\e^{\im\theta})^{k+2}f(q)\,\int_{\mathbb{C}_u\times S^2}\frac{\d u\wedge\D\lambda}{\la\lambda\hat{\lambda}\ra^2}\wedge h_{\bar\lambda}\wedge\bar\partial_u\left(\frac{\la\lambda\hat{\lambda}\ra^{k}\tilde\eta_{-2k}}{u^k}\right) 
\end{equation}
in order for the $S^1$ integral to survive, one is required to have $f(q)=q^{-k-2}$, which essentially isolates boundary radiative profile with specific dependence in bulk variable. We can further isolate the anti-holomorphic form degree from $h_{\bar\lambda}$ by decomposing it as $h_{\bar\lambda}=\eta\D\bar\lambda$, with $\eta$ signalling some negative helicity radiative degrees of freedom on $\scri$, whose $u$-derivative is usually referred to as the News tensor for asymptotic radiative degree of freedom. Such kinetic term suggests the symplectic pairing between $\eta$ and $\tilde\eta$ is given by the orthogonal relation
\begin{equation}
    \eta(\lambda_1,u_1)\tilde\eta(\lambda_2,u_2)\sim \frac{\delta(\la\lambda_1\lambda_2\ra)\delta([\bar\lambda_1\bar\lambda_2])}{u_1-u_2}
\end{equation}
Substituting $h_{\bar\lambda}$ in our large gauge profile, we have 
\begin{equation}
    Q_k=\int_{\mathbb{C}\times S^2} \D\lambda\wedge\d u \wedge \eta\D\bar\lambda\wedge\bar\partial_u\left(\frac{\la\lambda\hat{\lambda}\ra^k}{u^k}\,\tilde\eta_{-2k} \right)
\end{equation}
where for generic integers $k\geq 1$, the integral localises to the locus of $u=0$ with the large pure diffeomorphism generator allowed to be extended to higher orders in $u$ away from $u=0$:
\begin{equation}
    Q_k = \int_{S^2} \D\lambda\wedge\D\bar\lambda\,\partial^{k-1}_u\eta \tilde\eta_{-2k} \la\lambda\hat{\lambda}\ra^k
\end{equation}
As the charge of the large pure diffeomorphism generator $\tilde\eta$, we can check that $Q_k$ acts on $\tilde\eta$ as 
\begin{equation}
\begin{split}
        &Q_k\,\tilde\eta(\lambda_2,u_2)= \int \D\lambda_1\wedge\d u_1 \wedge \eta\D\bar\lambda_1\wedge\bar\partial_{u_1}\left(\frac{\la\lambda_1\hat{\lambda}_1\ra^k}{u_1^k}\,\tilde\eta(\lambda_1,u_1) \right)\tilde\eta(\lambda_2,u_2)\\
        \sim & \int \D\lambda_1\wedge\d u_1 \wedge \D\bar\lambda_1\wedge\bar\delta(u_1-u_2)\,\delta(\la\lambda_1\lambda_2\ra)\delta([\bar\lambda_1\bar\lambda_2])\,\frac{\la\lambda_1\hat{\lambda}_1\ra^k}{u_1^k}\,\tilde\eta(\lambda_1,u_1) \nonumber\\
        & = \frac{\tilde\eta(\lambda_2,u_2)}{u_2^k}
\end{split}
\end{equation}
which shifts $\tilde\eta$ by giving it extra singular behaviours along $\scri$.

\bibliographystyle{JHEP}
\bibliography{CCFT}

\providecommand{\href}[2]{#2}\begingroup\raggedright\begin{thebibliography}{10}

\bibitem{Kapec:2014zla}
D.~Kapec, V.~Lysov, and A.~Strominger, {\it {Asymptotic Symmetries of Massless QED in Even Dimensions}},  {\em Adv. Theor. Math. Phys.} {\bf 21} (2017) 1747--1767, [\href{http://arxiv.org/abs/1412.2763}{{\tt arXiv:1412.2763}}].

\bibitem{Gabai:2016kuf}
B.~Gabai and A.~Sever, {\it {Large gauge symmetries and asymptotic states in QED}},  {\em JHEP} {\bf 12} (2016) 095, [\href{http://arxiv.org/abs/1607.08599}{{\tt arXiv:1607.08599}}].

\bibitem{Campiglia:2016hvg}
M.~Campiglia and A.~Laddha, {\it {Subleading soft photons and large gauge transformations}},  {\em JHEP} {\bf 11} (2016) 012, [\href{http://arxiv.org/abs/1605.09677}{{\tt arXiv:1605.09677}}].

\bibitem{Campiglia:2016efb}
M.~Campiglia and A.~Laddha, {\it {Sub-subleading soft gravitons and large diffeomorphisms}},  {\em JHEP} {\bf 01} (2017) 036, [\href{http://arxiv.org/abs/1608.00685}{{\tt arXiv:1608.00685}}].

\bibitem{Strominger:2017zoo}
A.~Strominger, {\it {Lectures on the Infrared Structure of Gravity and Gauge Theory}},  \href{http://arxiv.org/abs/1703.05448}{{\tt arXiv:1703.05448}}.

\bibitem{Duary:2022onm}
S.~Duary, {\it {Celestial amplitude for 2d theory}},  {\em JHEP} {\bf 12} (2022) 060, [\href{http://arxiv.org/abs/2209.02776}{{\tt arXiv:2209.02776}}].

\bibitem{Donnay:2018neh}
L.~Donnay, A.~Puhm, and A.~Strominger, {\it {Conformally Soft Photons and Gravitons}},  {\em JHEP} {\bf 01} (2019) 184, [\href{http://arxiv.org/abs/1810.05219}{{\tt arXiv:1810.05219}}].

\bibitem{Pasterski:2021fjn}
S.~Pasterski, A.~Puhm, and E.~Trevisani, {\it {Celestial diamonds: conformal multiplets in celestial CFT}},  {\em JHEP} {\bf 11} (2021) 072, [\href{http://arxiv.org/abs/2105.03516}{{\tt arXiv:2105.03516}}].

\bibitem{Donnay:2020guq}
L.~Donnay, S.~Pasterski, and A.~Puhm, {\it {Asymptotic Symmetries and Celestial CFT}},  {\em JHEP} {\bf 09} (2020) 176, [\href{http://arxiv.org/abs/2005.08990}{{\tt arXiv:2005.08990}}].

\bibitem{Pasterski:2016qvg}
S.~Pasterski, S.-H. Shao, and A.~Strominger, {\it {Flat Space Amplitudes and Conformal Symmetry of the Celestial Sphere}},  {\em Phys. Rev. D} {\bf 96} (2017), no.~6 065026, [\href{http://arxiv.org/abs/1701.00049}{{\tt arXiv:1701.00049}}].

\bibitem{Pasterski:2017kqt}
S.~Pasterski and S.-H. Shao, {\it {Conformal basis for flat space amplitudes}},  {\em Phys. Rev. D} {\bf 96} (2017), no.~6 065022, [\href{http://arxiv.org/abs/1705.01027}{{\tt arXiv:1705.01027}}].

\bibitem{Bu:2023cef}
W.~Bu and S.~Seet, {\it {Celestial holography and AdS3/CFT2 from a scaling reduction of twistor space}},  \href{http://arxiv.org/abs/2306.11850}{{\tt arXiv:2306.11850}}.

\bibitem{Bu:2023vjt}
W.~Bu and S.~Seet, {\it {A hidden 2d CFT for self-dual Yang-Mills on the celestial sphere}},  \href{http://arxiv.org/abs/2310.17457}{{\tt arXiv:2310.17457}}.

\bibitem{Sleight:2023ojm}
C.~Sleight and M.~Taronna, {\it {Celestial Holography Revisited}},  \href{http://arxiv.org/abs/2301.01810}{{\tt arXiv:2301.01810}}.

\bibitem{Cheung:2016iub}
C.~Cheung, A.~de~la Fuente, and R.~Sundrum, {\it {4D scattering amplitudes and asymptotic symmetries from 2D CFT}},  {\em JHEP} {\bf 01} (2017) 112, [\href{http://arxiv.org/abs/1609.00732}{{\tt arXiv:1609.00732}}].

\bibitem{deBoer:2003vf}
J.~de~Boer and S.~N. Solodukhin, {\it {A Holographic reduction of Minkowski space-time}},  {\em Nucl. Phys. B} {\bf 665} (2003) 545--593, [\href{http://arxiv.org/abs/hep-th/0303006}{{\tt hep-th/0303006}}].

\bibitem{Kim:2023qbl}
S.~Kim, P.~Kraus, R.~Monten, and R.~M. Myers, {\it {S-matrix path integral approach to symmetries and soft theorems}},  {\em JHEP} {\bf 10} (2023) 036, [\href{http://arxiv.org/abs/2307.12368}{{\tt arXiv:2307.12368}}].

\bibitem{Strominger:2021mtt}
A.~Strominger, {\it {$w_{1+\infty}$ Algebra and the Celestial Sphere: Infinite Towers of Soft Graviton, Photon, and Gluon Symmetries}},  {\em Phys. Rev. Lett.} {\bf 127} (2021), no.~22 221601.

\bibitem{Guevara:2021abz}
A.~Guevara, E.~Himwich, M.~Pate, and A.~Strominger, {\it {Holographic symmetry algebras for gauge theory and gravity}},  {\em JHEP} {\bf 11} (2021) 152, [\href{http://arxiv.org/abs/2103.03961}{{\tt arXiv:2103.03961}}].

\bibitem{Costello:2022_CP}
K.~Costello and N.~M. Paquette, {\it {Celestial holography meets twisted holography: 4d amplitudes from chiral correlators}},  {\em JHEP} {\bf 10} (2022) 193, [\href{http://arxiv.org/abs/2201.02595}{{\tt arXiv:2201.02595}}].

\bibitem{Costello:2023hmi}
K.~Costello, N.~M. Paquette, and A.~Sharma, {\it {Burns space and holography}},  \href{http://arxiv.org/abs/2306.00940}{{\tt arXiv:2306.00940}}.

\bibitem{Adamo:2021lrv}
T.~Adamo, L.~Mason, and A.~Sharma, {\it {Celestial $w_{1+\infty}$ Symmetries from Twistor Space}},  {\em SIGMA} {\bf 18} (2022) 016, [\href{http://arxiv.org/abs/2110.06066}{{\tt arXiv:2110.06066}}].

\bibitem{Mason:2022hly}
L.~Mason, {\it {Gravity from holomorphic discs and celestial $Lw_{1+\infty}$ symmetries}},  \href{http://arxiv.org/abs/2212.10895}{{\tt arXiv:2212.10895}}.

\bibitem{Bittleston:2024rqe}
R.~Bittleston, G.~Bogna, S.~Heuveline, A.~Kmec, L.~Mason, and D.~Skinner, {\it {On AdS$_4$ deformations of celestial symmetries}},  \href{http://arxiv.org/abs/2403.18011}{{\tt arXiv:2403.18011}}.

\bibitem{Bern:1998sv}
Z.~Bern, L.~J. Dixon, M.~Perelstein, and J.~S. Rozowsky, {\it {Multileg one loop gravity amplitudes from gauge theory}},  {\em Nucl. Phys. B} {\bf 546} (1999) 423--479, [\href{http://arxiv.org/abs/hep-th/9811140}{{\tt hep-th/9811140}}].

\bibitem{Nguyen:2009jk}
D.~Nguyen, M.~Spradlin, A.~Volovich, and C.~Wen, {\it {The Tree Formula for MHV Graviton Amplitudes}},  {\em JHEP} {\bf 07} (2010) 045, [\href{http://arxiv.org/abs/0907.2276}{{\tt arXiv:0907.2276}}].

\bibitem{Adamo:2013tja}
T.~Adamo and L.~Mason, {\it {Conformal and Einstein gravity from twistor actions}},  {\em Class. Quant. Grav.} {\bf 31} (2014), no.~4 045014, [\href{http://arxiv.org/abs/1307.5043}{{\tt arXiv:1307.5043}}].

\bibitem{Adamo:2011cb}
T.~Adamo and L.~Mason, {\it {MHV diagrams in twistor space and the twistor action}},  {\em Phys. Rev. D} {\bf 86} (2012) 065019, [\href{http://arxiv.org/abs/1103.1352}{{\tt arXiv:1103.1352}}].

\bibitem{Penrose_Rindler_1988}
R.~Penrose and W.~Rindler, {\em Spinors and space-time, vol. 2.}
\newblock Cambridge University Press, 1988.

\bibitem{Chalmers:1996rq}
G.~Chalmers and W.~Siegel, {\it {The Selfdual sector of QCD amplitudes}},  {\em Phys. Rev. D} {\bf 54} (1996) 7628--7633, [\href{http://arxiv.org/abs/hep-th/9606061}{{\tt hep-th/9606061}}].

\bibitem{Capovilla:1991qb}
R.~Capovilla, T.~Jacobson, J.~Dell, and L.~J. Mason, {\it {Selfdual two forms and gravity}},  {\em Class. Quant. Grav.} {\bf 8} (1991) 41--57.

\bibitem{Urbantke_1984}
H.~Urbantke, {\it {On integrability properties of SU (2) Yang–Mills fields. I. Infinitesimal part}},  {\em Journal of Mathematical Physics} {\bf 25} (07, 1984) 2321--2324, [\href{http://arxiv.org/abs/https://pubs.aip.org/aip/jmp/article-pdf/25/7/2321/19167635/2321\_1\_online.pdf}{{\tt https://pubs.aip.org/aip/jmp/article-pdf/25/7/2321/19167635/2321\_1\_online.pdf}}].

\bibitem{Capovilla:1991kx}
R.~Capovilla, T.~Jacobson, and J.~Dell, {\it {A Pure spin connection formulation of gravity}},  {\em Class. Quant. Grav.} {\bf 8} (1991) 59--73.

\bibitem{Bittleston:2022nfr}
R.~Bittleston, A.~Sharma, and D.~Skinner, {\it {Quantizing the non-linear graviton}},  \href{http://arxiv.org/abs/2208.12701}{{\tt arXiv:2208.12701}}.

\bibitem{Mason:2007ct}
L.~J. Mason and M.~Wolf, {\it {Twistor Actions for Self-Dual Supergravities}},  {\em Commun. Math. Phys.} {\bf 288} (2009) 97--123, [\href{http://arxiv.org/abs/0706.1941}{{\tt arXiv:0706.1941}}].

\bibitem{Sharma:2021gcz}
A.~Sharma, {\it {Ambidextrous light transforms for celestial amplitudes}},  {\em JHEP} {\bf 01} (2022) 031, [\href{http://arxiv.org/abs/2107.06250}{{\tt arXiv:2107.06250}}].

\bibitem{Penrose:1985bww}
R.~Penrose and W.~Rindler, {\em {Spinors and Space-Time}}.
\newblock Cambridge Monographs on Mathematical Physics. Cambridge Univ. Press, Cambridge, UK, 4, 2011.

\bibitem{Penrose:1986ca}
R.~Penrose and W.~Rindler, {\em {SPINORS AND SPACE-TIME. VOL. 2: SPINOR AND TWISTOR METHODS IN SPACE-TIME GEOMETRY}}.
\newblock Cambridge Monographs on Mathematical Physics. Cambridge University Press, 4, 1988.

\bibitem{costello2021quantizing}
K.~J. Costello, {\it Quantizing local holomorphic field theories on twistor space},  2021.

\bibitem{Jones:1985pla}
P.~Jones and K.~Tod, {\it {Minitwistor spaces and Einstein-Weyl spaces}},  {\em Class. Quant. Grav.} {\bf 2} (1985), no.~4 565--577.

\bibitem{Calderbank98einstein-weylgeometry}
D.~M.~J. Calderbank and H.~Pedersen, {\it Einstein-weyl geometry},  {\em Adv. Math} {\bf 97} (1998) 74--109.

\bibitem{Hitchin:1982gh}
N.~J. Hitchin, {\it {MONOPOLES AND GEODESICS}},  {\em Commun. Math. Phys.} {\bf 83} (1982) 579--602.

\bibitem{Hitchin:1983ay}
N.~J. Hitchin, {\it {On the Construction of Monopoles}},  {\em Commun. Math. Phys.} {\bf 89} (1983) 145--190.

\bibitem{Donnay:2024qwq}
L.~Donnay, L.~Freidel, and Y.~Herfray, {\it {Carrollian $Lw_{1+\infty}$ representation from twistor space}},  \href{http://arxiv.org/abs/2402.00688}{{\tt arXiv:2402.00688}}.

\bibitem{Himwich:2020rro}
E.~Himwich, S.~A. Narayanan, M.~Pate, N.~Paul, and A.~Strominger, {\it {The Soft $\mathcal{S}$-Matrix in Gravity}},  {\em JHEP} {\bf 09} (2020) 129, [\href{http://arxiv.org/abs/2005.13433}{{\tt arXiv:2005.13433}}].

\bibitem{Kulish:1970ut}
P.~P. Kulish and L.~D. Faddeev, {\it {Asymptotic conditions and infrared divergences in quantum electrodynamics}},  {\em Theor. Math. Phys.} {\bf 4} (1970) 745.

\bibitem{Nguyen:2023ibj}
K.~Nguyen, A.~Rios~Fukelman, and C.~D. White, {\it {Celestial soft dressings from generalised Wilson lines}},  {\em PoS} {\bf CORFU2022} (2023) 289, [\href{http://arxiv.org/abs/2304.01250}{{\tt arXiv:2304.01250}}].

\bibitem{Hodges:2012ym}
A.~Hodges, {\it {A simple formula for gravitational MHV amplitudes}},  \href{http://arxiv.org/abs/1204.1930}{{\tt arXiv:1204.1930}}.

\bibitem{He:2024ddb}
T.~He, P.~Mitra, A.~Sivaramakrishnan, and K.~M. Zurek, {\it {An On-Shell Derivation of the Soft Effective Action in Abelian Gauge Theories}},  \href{http://arxiv.org/abs/2403.14502}{{\tt arXiv:2403.14502}}.

\bibitem{Kapec:2021eug}
D.~Kapec and P.~Mitra, {\it {Shadows and soft exchange in celestial CFT}},  {\em Phys. Rev. D} {\bf 105} (2022), no.~2 026009, [\href{http://arxiv.org/abs/2109.00073}{{\tt arXiv:2109.00073}}].

\bibitem{LeBrun1991ExplicitSM}
C.~LeBrun, {\it Explicit self-dual metrics on cp 2 \# \#cp 2},  1991.

\bibitem{Adamo:2014yya}
T.~Adamo, E.~Casali, and D.~Skinner, {\it {Perturbative gravity at null infinity}},  {\em Class. Quant. Grav.} {\bf 31} (2014), no.~22 225008, [\href{http://arxiv.org/abs/1405.5122}{{\tt arXiv:1405.5122}}].

\bibitem{Adamo:2021bej}
T.~Adamo, L.~Mason, and A.~Sharma, {\it {Twistor sigma models for quaternionic geometry and graviton scattering}},  \href{http://arxiv.org/abs/2103.16984}{{\tt arXiv:2103.16984}}.

\bibitem{Geyer:2014fka}
Y.~Geyer, A.~E. Lipstein, and L.~J. Mason, {\it {Ambitwistor Strings in Four Dimensions}},  {\em Phys. Rev. Lett.} {\bf 113} (2014), no.~8 081602, [\href{http://arxiv.org/abs/1404.6219}{{\tt arXiv:1404.6219}}].

\bibitem{Caron-Huot:2023wdh}
S.~Caron-Huot, F.~Coronado, and B.~M\"uhlmann, {\it {Determinants in self-dual $ \mathcal{N} $ = 4 SYM and twistor space}},  {\em JHEP} {\bf 08} (2023) 008, [\href{http://arxiv.org/abs/2304.12341}{{\tt arXiv:2304.12341}}].

\end{thebibliography}\endgroup

\end{document}